\documentclass[a4paper,11pt]{article}	

\usepackage{classeJB}

\usepackage{arydshln}
\usepackage{algorithm}
\usepackage[noend]{algpseudocode}

\newcommand*\unit[1]{\bigl[\, \mathsf{#1} \,\bigr]}

\newcommand{\apr}{\mathrm{apr}}

\newcommand*\prob[1]{\pi \,\bigl(\, #1 \,\bigr)}
\newcommand*\p{\boldsymbol{p}}

\title{
\vspace{-1.5cm}
Parameter estimation and model selection for water sorption in a wood fibre material
\vspace{4pt}
}

\author{Julien Berger \textsuperscript{a}$^{\ast}$, Thibaut Colinart\textsuperscript{b}, Bruna R. Loiola\textsuperscript{c}, Helcio R.B. Orlande\textsuperscript{d}\\
\date{\today\vspace{-0.5cm}}}

\begin{document}

\maketitle

\begin{center}
\small
\textsuperscript{a} Laboratoire des Sciences de l’Ingénieur pour l’Environnement (LaSIE), UMR 7356 CNRS, La Rochelle Université, CNRS, 17000, La Rochelle, France \\
\textsuperscript{b} Univ. Bretagne Sud, UMR CNRS 6027, IRDL, 56100 Lorient, France\\
\textsuperscript{c} Military Institute of Engineering, Mechanical Engineering Department, Rio de Janeiro
- 22290-270, Brazil \\
\textsuperscript{d} POLI/COPPE, Mechanical Engineering Graduate Program, Federal University of Rio de Janeiro \\
$^{\ast}$corresponding author, e-mail address : julien.berger@univ-lr.fr\\
\end{center}

\begin{abstract}

The sorption curve is an essential feature for the modelling of heat and mass transfer in porous building materials. Several models have been proposed in the literature to represent the amount of moisture content in the material according to the water activity (or capillary pressure) level. These models are based on analytical expressions and few parameters that need to be estimated by inverse analysis. This article investigates the reliability of eight models through the accuracy of the estimated parameters. For this, experimental data for a wood fibre material are generated with special attention to the stop criterion to capture long time kinetic constants. Among five sets of measurements, the best estimate is computed. The reliability of the models is then discussed. After proving the theoretical identifiability of the unknown parameters for each model, the primary identifiability is analysed. It evaluates whether the parameters influence on the model output is sufficient to proceed the parameter estimation with accuracy. For this, a continuous derivative-based approach is adopted. Seven models have a low primary identifiability for at least one parameter. Indeed, when estimating the unknown parameters using the experimental observations, the parameters with low primary identifiability exhibit large uncertainties. Finally, an Approximation Bayesian Computation algorithm is used to simultaneously select the best model and estimate the parameters that best represent the experimental data. The thermodynamic and \textsc{Feng}--\textsc{Xing} models, together with a proposed model in this work, were the best ones selected by this algorithm.

\textbf{Keywords:} sorption models in building porous materials; model reliability; parameter estimation problem; primary identifiability; ABC algorithm

\end{abstract}

\section{Introduction}

Within the environmental context, bio-based materials such as wood fiber have been increasingly used in building constructions due to their reduced ecological footprint and their thermal performance. The properties of such wood-based materials have a good reproducibility due to their industrial production. Several recent studies pointed out the importance of modeling accurately the phenomena of adsorption in such materials to predict the phenomena of heat and mass transfer and assess the moisture disorder risks \citep{berger_2015}. In the \emph{Conclusion} Section of \citep{Patera_2016}, the error in the prediction can reach $20$ to $30\%$ of moisture content in the wood without modeling accurately the adsorption phenomena combined with hysteresis effects.  Those results are confirmed by \citep{Zhang_2016b} for a wood fiber material where the authors highlight the importance of modeling accurately adsorption phenomena to evaluate the risk of mold growth. Furthermore, the moisture content is a crucial parameter in the modeling framework since several other properties such as thermal conductivity depend on it \citep{Willems_2014}.

As presented in \citep{Skaar_1988}, several phenomenological models have been proposed in the literature to reproduce the moisture sorption curve. Among them, one can mention the \textsc{Guggenheim}--\textsc{Anderson}--\textsc{de Boer} (GAB) model with examples of applications in \citep{Singh_1996,Iglesias_1995}. The \textsc{Brunauer}, \textsc{Emmett} and \textsc{Teller} model is employed in \citep{Colinart_2017} for a hygroscopic material. In \citep{Carmeliet_2002}, the \textsc{van Genuchten} model is used for several building materials. All these models are based on a few important parameters that can be estimated using experimental data of moisture content according to the water activity or capillary pressure. These experimental observations can be obtained using static gravimetric methods. Then, the unknown parameters can be inferred by solving an inverse problem. For instance, the parameters of several models are retrieved in \citep{Iglesias_1995}, \citep{Ouertani_2014} or \citep{Stolarska_2017}. Those studies use least square estimator algorithm and the discussion among the models is based only on the residual between numerical predictions and experimental data. In \citep{Furmaniak_2012,Karoglou_2005} investigations are carried considering one model. The parameters are determined for several materials without information on the accuracy of the estimation.

It is essential to have reliable models, which accurately represent the physical phenomena when compared to experimental observations. Such evaluation has already been done for wood materials in \citep{Zhang_2015}. Nevertheless, the reliability is also based on the capacity of estimating the important parameters with accuracy. This accurate estimation is also required to verify the theoretical models based on first principles of physics. This paper proposes to investigate the reliability of eight models, seven being mainly used for heat and mass transfer modelling in building materials. The last one is proposed according to the general curve of moisture sorption. To discuss the reliability, experimental measurements in a wood fibre material are taken using a DVS equipment. Then, using the experimental observations, the parameter estimation problem can be solved for each model. The accuracy of the retrieved parameters and the resultant robustness of the models are discussed using two important approaches. First, the primary identifiability of the parameters is evaluated. It enables to evaluate the sensitivity of the output models to each unknown parameter, using a continuous derivative-based approach. If a model is not sensitive to a parameter, it indicates that the latter cannot be retrieved with accuracy. Then, an efficient Approximate Bayesian Computation (ABC) algorithm is employed to conduct a selection over the eight competing models. The selection is achieved sequentially with decreasing tolerances. The selected model is the one having the highest probability to minimize the distance between predictions and experimental observations for the smallest tolerance. 

The article is organized as follows. Section~\ref{sec:physical_model} presents the eight physical models to predict the moisture content according to the water activity. Moreover, the methodology is described to evaluate the robustness of the models in the framework of a parameter estimation problem. Section~\ref{sec:experimental_measurement} introduces the experimental measurements obtained for a wood fibre material. Then, Section~\ref{sec:reliability_models} discusses the reliability of the models and Section~\ref{sec:conclusion} gives some general remarks on the results of the study.

\section{Physical model}
\label{sec:physical_model}

\subsection{Models for water adsorption}

The sorption model of water in porous material describes the water content $u \ \unit{-}$ contained in the porous matrix for a defined water activity $a \ \unit{-}\,$. The moisture content $u$ is defined as
\begin{align*}
\displaystyle u \ \eqdef \frac{m \moins m_{\,0}}{m_{\,0}}\,,
\end{align*} 
where $m_{\,0} \ \unit{kg}$ is the dry mass of the material and $m \ \unit{kg}$ the mass of the material. In the literature, several models are proposed to represent the dependency of $u$ on the water activity $a\,$. 
From a mathematical point of view, the sorption curve can be formulated as:
\begin{align*}
u \,:\, \bigcup_{n=1}^N \Omega_{\,p_{\,n}} \times \Omega_{\,a} & \longrightarrow \, \mathbb{R}_{\,>0\,}  \,, \\
\bigl(\,p_{\,1}\,,\, \ldots \,,\, p_{\,N} \,,\,a\,\bigr) & \longmapsto \, f\;\bigl(\,p_{\,1}\,,\, \ldots \,,\, p_{\,N} \,,\,a\,\bigr)  \,,
\end{align*}
where $a$ is the water activity, $f \ \unit{-} $ is the sorption model and $p_{\,n}\,, \quad \forall n \, \in \, \bigl\{\, 1 \,,\, \ldots \,,\, N\,\bigr\}$ is an unknown parameter involved in the model definition. Depending on the model, the parameter $p_{\,n}$ may have a physical meaning and specific unit.
As presented below, the total number of parameter $N$ varies from two to four depending on the sorption model. The set  of the parameter $p_{\,n}$ verifies $\Omega_{\,p_{\,n}} \, \subset \, \mathbb{R}\,$. The water activity $a$ belongs to the interval $\bigl[\,0\,,\,1\,\bigr]\,$. However, $a$ is restricted to $\Omega_{\,a} \egal \bigl[\,0.05\,,\,0.95\,\bigr]$ for several reasons. First, some models proposed in the literature are \emph{mathematically} not defined for $a \egal 0\,$. Then, in this study, it is stated that the moisture content at saturation $a\egal 1$ is unknown. Indeed, for this value, moisture content may cover several definitions like hygroscopic, capillary or complete saturation moisture content \citep{Nilsson_2018}. Moreover, the discussion is carried out with the perspective of using sorption models for the simulation of heat and mass transfer in porous building materials under normal conditions. Thus, the fully dry and saturated states are never reached in practice. 

A total of eight models are investigated, namely the \textsc{Brunauer}, \textsc{Emmett} and \textsc{Teller} (BET), the
normalized \textsc{Guggenheim}--\textsc{Anderson}--\textsc{de Boer} (GAB), the thermodynamic (TRM), the empirical \textsc{Oswin} (OSW), the \textsc{Feng}--\textsc{Xing} (FX), the \textsc{van Genuchten} (VG) and the \textsc{Smith} (SM) one. An additional  moisture adsorption (MADS) model is proposed based on the general shape of the sorption curve. A detailed presentation of each model is now given. The indicator $m \, \in \, \bigl\{\, 1 \,,\, \ldots \,,\, 8 \,\bigr\}$ is linked to each model $u_{\,m} \ \equiv \ f_{\,m}$.

First, the so-called \textsc{Brunauer}, \textsc{Emmett} and \textsc{Teller} (BET) model \citep{Brunauer_1938} is defined by:
\begin{align}
\label{eq:BET}
f_{\,1}\;\bigl(\,p_{\,1}\,,\, p_{\,2} \,,\,a\,\bigr) \egal 
\frac{ p_{\,1} \ p_{\,2}}{\Bigl(\,1 \moins a \,\Bigr) \ \Bigl(\, 1 \plus \bigl(\, p_{\,2}\moins 1 \,\bigr) \ a\,\Bigr)} \cdot a \,,
\end{align}
which is not valid for $a \egal 1\,$. According to \citep{Blahovec_2008}, the model fits well experimental data for a water activity from $0$ to $\mathcal{O}(\,0.6\,)\,$. 

The second model is the normalized \textsc{Guggenheim}--\textsc{Anderson}--\textsc{de Boer} (GAB) model:
\begin{align}
\label{eq:GAB_model}
f_{\,2}\;\bigl(\,p_{\,1}\,,\, p_{\,2} \,,\, p_{\,3} \,,\,a\,\bigr) \egal 
\frac{ p_{\,1} \ p_{\,2} \ p_{\,3}}{\bigl(\,1 \moins p_{\,2} \ a \,\bigr) \, \bigl(\, 1 \moins p_{\,2} \ a \plus p_{\,2} \ p_{\,3} \ a\,\bigr)} \ a \,.
\end{align}
The parameter $p_{\,1}$ represents the normalized moisture content to monomolecular layer. Investigations from \citep{Blahovec_2004} suggest that the model may represent well the phenomena for a water activity lower than $\mathcal{O}(\,0.9\,)\,$.
It can be remarked that the \textsc{Hailwood} model \citep{Hailwood_1946} is not investigated here due to its equivalence with the GAB model. 

The thermodynamic (TRM) model is given by \citep{Merakeb_2009}:
\begin{align}
\label{eq:TRM_model}
f_{\,3}\;\bigl(\,p_{\,1}\,,\, p_{\,2} \,,\,p_{\,3} \,,\,a\,\bigr) \egal 
p_{\,1} \ \exp \Bigl(\, p_{\,2} \ \ln \bigl(\, a \,\bigr) \ \exp \bigl(\, p_{\,3} \ a \,\bigr) \,\Bigr) \,.
\end{align}
It can be noticed that this model is not defined for $a \egal 0\,$, justifying again the definition of $\Omega_{\,a}\,$. According to \citep{Merakeb_2009}, the model is reliable for the whole domain of water activity. 

The empirical \textsc{Oswin} (OSW) model is defined as \citep{Oswin_1946,Ouertani_2014}:
\begin{align}
\label{eq:OSW_model}
f_{\,4}\;\bigl(\,p_{\,1}\,,\, p_{\,2} \,,\,a\,\bigr) \egal 
p_{\,1} \cdot \biggl(\, \frac{a}{1 \moins a} \,\biggr)^{\,p_{\,2}} \,.
\end{align} 
Here, the model is not defined for $a \egal 1\,$, giving another justification of the restriction of the range for the water activity $a\,$. 

The \textsc{Feng}--\textsc{Xing} (FX) model is \citep{Fredlund_1994}:
\begin{align*}
f_{\,5}\;\bigl(\,p_{\,1}\,,\, \tilde{p}_{\,2}\,,\, p_{\,3} \,,\, p_{\,4}\,,\,a\,\bigr) \egal p_{\,1} \cdot 
\Biggl[\, \ln \Biggl(\,\mathrm{e} 
\plus \biggl(\, \frac{\Psi}{\tilde{p}_{\,2}} \,\biggr)^{\,p_{\,3}} \,\Biggr)
\,\Biggr]^{\,-\,p_{\,4}}
 \,,
\end{align*}
where $\Psi \ \unit{Pa}$ is the capillary pressure, related to the water activity $a$ according to:
\begin{align}
\label{eq:capillary_pressure}
\Psi \egal - \rho_{\,2} \ R_{\,1} \ T \ \ln \bigl(\, a \, \bigr) \,,
\end{align}
with $\rho_{\,2} \egal 10^{\,3} \ \mathsf{kg\,.\,m^{\,-3}}$ is the liquid water density, $R_{\,1} \egal 462 \ \mathsf{J\,.\,kg^{\,-1}\,.\,K^{\,-1}}$ is the water vapour gas constant and $T \egal 296.15 \ \mathsf{K}$ ($23 \ \mathsf{^{\,\circ}C}$) is the temperature. Thus, the FX model can be reformulated as:
\begin{align*}
f_{\,5}\;\bigl(\,p_{\,1}\,,\, \tilde{p}_{\,2}\,,\, p_{\,3} \,,\, p_{\,4}\,,\,a\,\bigr) \egal p_{\,1} \cdot 
\Biggl[\, \ln \Biggl(\,\mathrm{e} 
\plus \biggl(\, \frac{- \rho_{\,2}  \ R_{\,1} \ T}{\tilde{p}_{\,2}} \ \ln (\,a\,)\,\biggr)^{\,p_{\,3}} \,\Biggr)
\,\Biggr]^{\,-\,p_{\,4}}
 \,.
\end{align*}
The model is not defined for $a \egal 0\,$. Furthermore, it can be noticed that $\rho_{\,2}  \ R_{\,1} \ T \egal \mathcal{O}(\,10^{\,8}\,)$ and that $\tilde{p}_{\,2} \egal \mathcal{O}(\,10^{\,7}\,)$ according to \citep{Zhang_2015}. Thus, to avoid rounding errors and computational difficulties when solving the parameter estimation problem, the parameter $\tilde{p}_{\,2}$ is replaced by $p_{\,2} \, \hookleftarrow \, \displaystyle  \frac{\rho_{\,2}  \ R_{\,1} \ T}{\tilde{p}_{\,2}} \,$. The FX model is finally written as:
\begin{align*}
f_{\,5}\;\bigl(\,p_{\,1}\,,\, p_{\,2}\,,\, p_{\,3} \,,\, p_{\,4}\,,\,a\,\bigr) \egal p_{\,1} \cdot 
\Biggl[\, \ln \Biggl(\,\mathrm{e} 
\plus \biggl(\, - \, p_{\,2} \ \ln (\,a\,)\,\biggr)^{\,p_{\,3}} \,\Biggr)
\,\Biggr]^{\,-\,p_{\,4}}
 \,.
\end{align*}

The \textsc{van Genuchten} (VG) model with the \textsc{Mualem} approach is given by \citep{Van_Genuchten_1980}:
\begin{align*}
f_{\,6}\;\bigl(\,p_{\,1}\,,\, \breve{p}_{\,2}\,,\, p_{\,3}\,,\,a\,\bigr) \egal p_{\,1}  \cdot
\biggl( \, 1 \plus \Bigl(\, \breve{p}_{\,2} \ \Psi \,\Bigr)^{\,p_{\,3}} \,\biggr)^{\,-1\plus \frac{1}{p_{\,3}}}
 \,,
\end{align*}
Using equation~\eqref{eq:capillary_pressure} of the capillary pressure, we obtain:
\begin{align*}
f_{\,6}\;\bigl(\,p_{\,1}\,,\, \breve{p}_{\,2}\,,\, p_{\,3}\,,\,a\,\bigr) \egal p_{\,1}  \cdot
\biggl( \, 1 \plus \Bigl(\, - \, \breve{p}_{\,2} \ \rho_{\,2} \ R_{\,1} \ T \ \ln \bigl(\, a \, \bigr)  \,\Bigr)^{\,p_{\,3}} \,\biggr)^{\,-1\plus \frac{1}{p_{\,3}}}
 \,.
\end{align*}
Given the magnitude of $\rho_{\,2}  \ R_{\,1} \ T$ and that $\breve{p}_{\,2} \egal \mathcal{O}(\,10^{\,-8}\,)$ according to \citep{Zhang_2015}, the parameter $\breve{p}_{\,2}$ is replaced by $p_{\,2} \, \hookleftarrow \, \displaystyle  \rho_{\,2}  \ R_{\,1} \ T \ \breve{p}_{\,2} \, $. So the VG model is finally formulated as:
\begin{align*}
f_{\,6}\;\bigl(\,p_{\,1}\,,\, p_{\,2}\,,\, p_{\,3}\,,\,a\,\bigr) \egal p_{\,1} \cdot
\biggl( \, 1 \plus \Bigl(\, - \, p_{\,2} \cdot \ln \bigl(\, a \, \bigr)  \,\Bigr)^{\,p_{\,3}} \,\biggr)^{\,-1\plus \frac{1}{p_{\,3}}}
 \,.
\end{align*}
This model is also not defined for $a \egal 0\,$. Both FX and VG model have been defined for the capillary domain so that their accuracy for lower water activity require analysis.


The last model from the literature has been proposed by \textsc{Smith} (SM) \citep{Smith_1947} based on empirical concept:
\begin{align*}
f_{\,7}\;\bigl(\,p_{\,1}\,,\, p_{\,2}\,,\, a\,\bigr) \egal p_{\,1} \plus p_{\,2} \ \ln \bigl(\, 1 \moins a \,\bigr)
 \,,
\end{align*}
with only two parameters to be estimated. 

\begin{figure}
\centering
\subfigure[\label{fig:f_S} ]{\includegraphics[width=.45\textwidth]{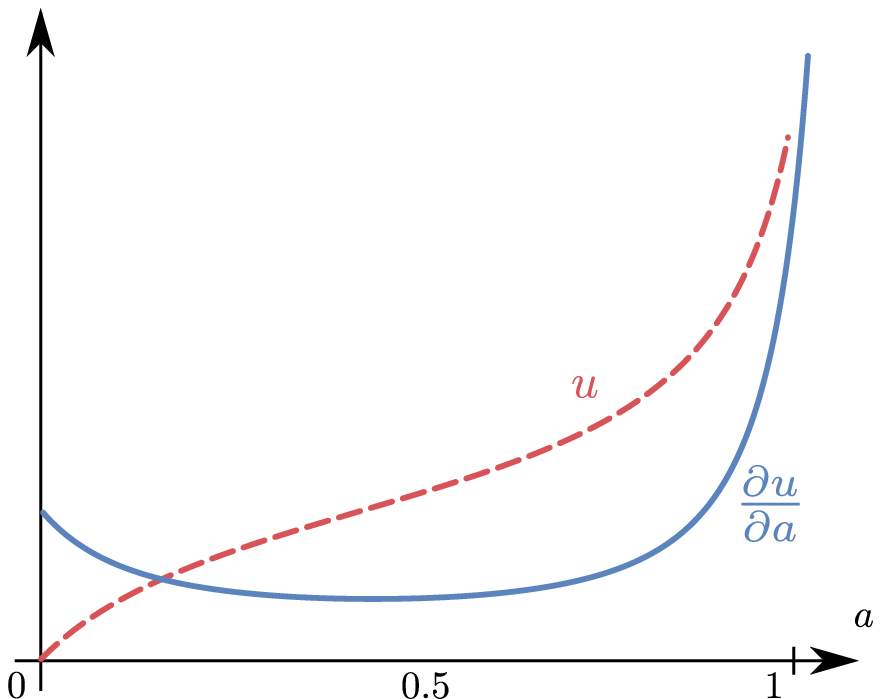}} \hspace{0.2cm}
\subfigure[\label{fig:f_tan}]{\includegraphics[width=.45\textwidth]{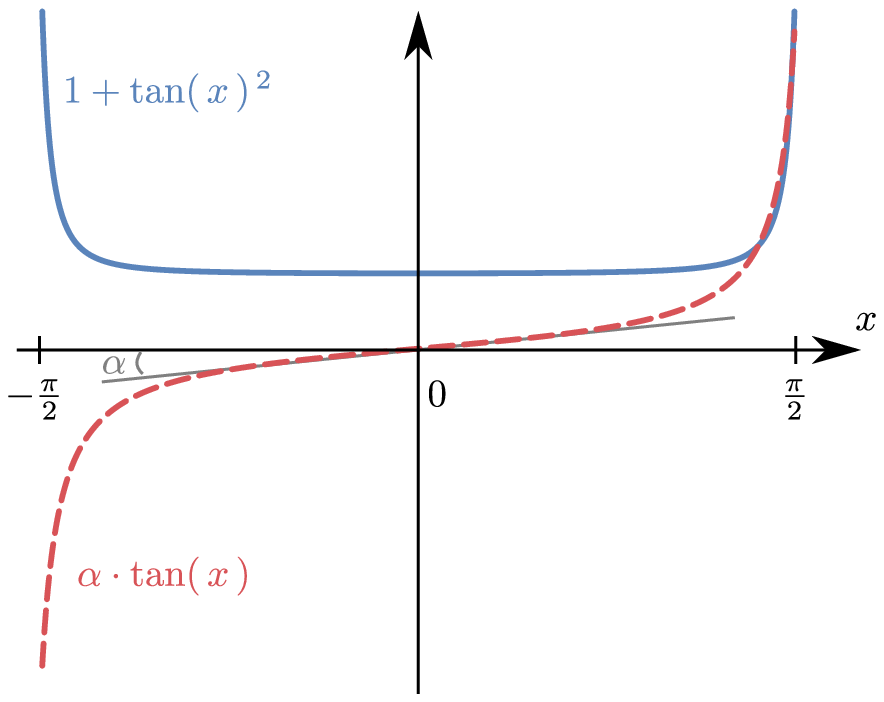}} 
\caption{Relation between the sorption curve for building porous materials and the mathematical functions.}
\label{fig:inspiration_model}
\end{figure}

A new model is proposed in this work to represent the physical phenomena. The general expression is the following:
\begin{align}
\label{eq:mads}
f_{\,8} \;\bigl(\,p_{\,1}\,,\, \ldots \,,\, p_{\,N} \,,\,a\,\bigr) \egal u_{\,0} \plus \alpha \ \tan \Bigl(\, P_{\,N}\,(\,a\,)\,\Bigr)  \,,
\end{align}
where $P_{\,N}\,(\,a\,)$ is a polynomial of order $N\,$, defined for $a \, \in \, \bigl[\, 0 \,,\, 1\,\bigr]\,$. Thus, the number of unknown parameters corresponds to the order of polynomial.
This proposal is inspired by the shape of the isotherm sorption for building porous material illustrated in Figure~\ref{fig:inspiration_model} in comparison to the general function $\alpha \, \tan (\,x\,)$ and its derivative $ \forall x \, \in \, \bigl]\, - \frac{\pi}{2} \,,\, \frac{\pi}{2}\,\bigr[\,$. Thus, the term $u_{\,0}$ in Eq.~\eqref{eq:mads} enables to switch the image of the model for positive values. The term $\alpha$ controls the slope of the model for middle value of water activity $a \egal 0.5\,$. Now, we consider the derivative of the model Eq.~\eqref{eq:mads} relatively to $a\,$:
\begin{align}
\label{eq:mads_derivee}
\pd{f_{\,8}}{a} \egal \alpha \ P^{\,\prime}_{\,N}(\,a\,) \cdot \biggl(\, 1 \plus \tan \Bigl(\, P_{\,N}\,(\,a\,)\,\Bigr)^{\,2} \,\biggr)  \,.
\end{align}
It can be remarked that the polynomial permits to regulate the positive slope of the sorption model. Indeed, in Figure~\ref{fig:f_S}, it can be remarked that the slope of $\displaystyle \pd{u}{a}$ has a specific shape. For $a \egal 0\,$, it has a certain value. When $a$  increases, the slope decreases until an almost constant value. Then, when $a \, > \, 0.85\,$, we reach the capillary state and the slopes increase exponentially. With this in mind, the parameters $u_{\,0}$ and $\alpha$ of the model are defined using the two following constraints:
\begin{subequations}
\label{eq:mads_const}
\begin{align}
f_{\,8} \;\bigl(\,p_{\,1}\,,\, \ldots \,,\, p_{\,N} \,,\,a \egal 0\,\bigr) & \egal 0 \,, \label{eq:mads_const1} \\[4pt]
\pd{f_{\,8}}{a} \;\bigl(\,p_{\,1}\,,\, \ldots \,,\, p_{\,N} \,,\,a \egal a_{\,0}\,\bigr) & \egal K \,, \label{eq:mads_const2}
\end{align}
\end{subequations}
where $K \ \unit{-}$ is the slope of the sorption curve at a given water activity $a_{\,0}\,$:
\begin{align*}
K \egal \displaystyle \pd{u}{a}\biggl|_{\,a\egal a_{\,0}} \,.
\end{align*} 
Thus, Eq.~\eqref{eq:mads_const1} enables to switch the image of the model for positive values. The second equation~\eqref{eq:mads_const2} forces the slope of the sorption model with a given value.
Here, the polynomial $P_{\,N} \egal P_{\,2}$ is assumed of first degree:
\begin{align*}
P_{\,2}\,(\,a\,) \egal p_{\,1} \plus p_{\,2} \ a \,.
\end{align*}
Using the constraints from Eq.~\eqref{eq:mads_const} and $a_{\,0} \egal 0\,$, the proposed model has the following formulation:
\begin{align}
\label{eq:MADS_model}
f_{\,8} \;\bigl(\,p_{\,1}\,,\,  p_{\,2} \,,\,a\,\bigr) \egal 
\frac{K}
{p_{\,2} \cdot \Bigl(\, 1 \plus \tan (\,p_{\,1}\,)^{\,2}\,\Bigr)} \cdot \Bigl(\, \tan (\, p_{\,1} \plus p_{\,2} \ a \,) \moins \tan (\, p_{\,1} \,) \,\Bigr) \,,
\end{align}
with two unknown parameters to be estimated. Equation~\eqref{eq:MADS_model} will be denoted as the Moisture ADSorption model (MADS). In the end, four models depend on two parameters, three models need three parameters and one model have four parameters to be determined.

\subsection{Parameter estimation problem}

The eight models depend on several parameters that can be determined using experimental observations of moisture content $u$ according to the water activity $a\,$. The procedure to solve the parameter estimation problem is now detailed. First, some important notations are clarified. It is assumed that measurement, denoted as $\hat{u}\,$, of moisture content in a material is obtained for different levels of water activity $\hat{a}\,$. The set of model indicator is:
\begin{align*}
\Omega_{\,m} \, \eqdef \, \Bigl\{\, 
m 
\,\Bigl|\,
m \, \in \, \bigl\{\, 1 \,,\, \ldots \,,\, 8 \,\bigr\}
 \,\Bigr\} \,,
\end{align*} 
where $m$ is an indicator representing one of the eight investigated models.
We denote as $u_{\,m} \, \eqdef \, f_{\,m}\,\bigl(\,\p \,,\,\hat{a}\,\bigr)$ the computed moisture content with the model $m$ for the water activity $\hat{a}$ of the experiments (with its respective domain $\Omega_{\,\hat{a}}$). In addition, for model $m\,$, the set of unknown parameters is defined as:
\begin{align*}
\Omega_{\,p\,,\,m} \, \eqdef \, \Bigl\{\, 
\p \egal \bigl(\,p_{\,1} \,,\,\ldots \,,\, p_{\,N} \bigr) 
\,\Bigl|\, u_{\,m} \egal f_{\,m}\,\bigl(\,\p \,,\,\hat{a}\,\bigr)
 \,\Bigr\}  \,.
\end{align*}
Several distinctions are made among the unknown parameters. First, the solution of the parameter estimation problem is denoted by $\p^{\,\circ}\,$. Then, the so-called \emph{a priori} parameters, which are used in the preliminary identifiability investigations, are written as $\p^{\,\apr}\,$. 

\subsubsection{Primary identifiability}

The primary identifiability of the unknown parameters is discussed according to \citep{Jumabekova_2019}. It aims at carrying a sensitivity analysis of the unknown parameters on the output model using a continuous derivative-based approach. For this, the sensitivity function $\theta_{\,n}$ of the model $f_{\,m}$ relatively to the parameter $p_{\,n}$ is defined \citep{Saltelli_2004}:
\begin{align*}
\theta_{\,n} \,:\,  \bigcup_{n=1}^N \Omega_{\,p_{\,n}}  \times \Omega_{\,a} & \longrightarrow \, \mathbb{R}  \,, \\
\bigl(\,p_{\,1}\,,\, \ldots \,,\, p_{\,N} \,,\,a\,\bigr) & \longmapsto \, \pd{f_{\,m}}{p_{\,n}} \,, \qquad \forall \, n \, \in \, \bigl\{\, 1 \,,\, \ldots \,,\, N\,\bigr\}  \,.
\end{align*}
The differentiation of sorption model relatively to the parameters is performed analytically. Then, the following sensitivity metrics are defined \citep{Sobol_1990,Dickinson_1976,Sobol_2009,Kucherenko_2016}:
\begin{align*}
\nu_{\,n}  \,:\, \bigcup_{n=1}^N \Omega_{\,p_{\,n}} \times \Omega_{\,a} & \longrightarrow \, \mathbb{R}  \,, \\
\bigl(\,p_{\,1}\,,\, \ldots \,,\, p_{\,N} \,,\,a\,\bigr) & \longmapsto \, 
\int_{\,\Omega_{\,p_{\,n}}} \, \theta_{\,n}^{\,2} \ \mathrm{d}p_{\,n}
 \,, \qquad \forall \, n \, \in \, \bigl\{\, 1 \,,\, \ldots \,,\, N\,\bigr\}  \,.
\end{align*}
and
\begin{align*}
\nu^{\,\intercal}_{\,n}  \,:\,\bigcup_{n=1}^N \Omega_{\,p_{\,n}}  \times \Omega_{\,a} & \longrightarrow \, \mathbb{R}  \,, \\
\bigl(\,p_{\,1}\,,\, \ldots \,,\, p_{\,N} \,,\,a\,\bigr) & \longmapsto \, 
\int_{\,\Omega_{\,a}} \; \int_{\,\Omega_{\,p}} \, \theta_{\,n}^{\,2} \, \mathrm{d}a \, \mathrm{d}p
 \,, \qquad \forall \, n \, \in \, \bigl\{\, 1 \,,\, \ldots \,,\, N\,\bigr\}  \,.
\end{align*}
The quantities $\nu_{\,n}$ and $\nu^{\,\intercal}_{\,n}$ translate how changes of parameter $p_{\,n}$ impact the sorption model $f_{\,m}\,$. The first one is local and depends on the value of the water activity $a\,$, while the second is global over the whole range of $a\,$. A large value of those metrics reveals an important influence of the parameter. For the analysis, it is transformed into a dimensionless metric to get the derivative-based sensitivity indexes $\gamma_{\,n}$ and $\gamma^{\,\intercal}_{\,n} \,$:
\begin{align*}
\gamma_{\,n}  \,:\, \bigcup_{n=1}^N \Omega_{\,p_{\,n}}  \times \Omega_{\,a} & \longrightarrow \, \mathbb{R}  \,, \\
\bigl(\,p_{\,1}\,,\, \ldots \,,\, p_{\,N} \,,\,a\,\bigr) & \longmapsto \, 
\frac{\nu_{\,n}}{\displaystyle \sum_{n=1}^{N} \, \nu_{\,n} }
 \,, \qquad \forall n \, \in \, \bigl\{\, 1 \,,\, \ldots \,,\, N\,\bigr\}  \,.
\end{align*}
and 
\begin{align*}
\gamma^{\,\intercal}_{\,n}  \,:\, \bigcup_{n=1}^N \Omega_{\,p_{\,n}}  \times \Omega_{\,a} & \longrightarrow \, \mathbb{R}  \,, \\
\bigl(\,p_{\,1}\,,\, \ldots \,,\, p_{\,N} \,,\,a\,\bigr) & \longmapsto \, 
\frac{\nu^{\,\intercal}_{\,n}}{\displaystyle \sum_{n=1}^{N} \, \nu^{\,\intercal}_{\,n} }
 \,, \qquad \forall n \, \in \, \bigl\{\, 1 \,,\, \ldots \,,\, N\,\bigr\}  \,.
\end{align*}
Both metrics $\nu_{\,n}$ and $\gamma_{\,n}$ assess the sensitivity of the parameter over its whole domain of variation $\Omega_{\,p_{\,n}}\,$. The metrics $\gamma_{\,n}$ and $\gamma^{\,\intercal}_{\,n} $ are local and global according to the water activity, respectively.

With the computation of the sensitivity functions, the so-called \textsc{Fisher} information matrix $F$ \citep{Karalashvili_2015,Ucinski} is defined as:
\begin{align*}
 F \,:\, \bigcup_{n=1}^N \Omega_{\,p_{\,n}}  \times \Omega_{\,a} & \longrightarrow \, \mathcal{M}\bigl(\, \mathbb{R}^{\,N \times N} \,\bigr)  \,, \\
 \bigl(\,p_{\,1}\,,\, \ldots \,,\, p_{\,N} \,,\,a\,\bigr) & \longmapsto \, 
 \bigl[\, F_{\,n_{\,1} \, n_{\,2}} \,\bigr] \,, && \forall \, (\,n_{\,1}\,,\,n_{\,2}\,) \, \in \, \bigl\{ 1 \,,\, \ldots \,,\, N \, \bigr\}^{\,2}\,, \\[3pt]
 F_{\,n_{\,1} \, n_{\,2}} & \, \eqdef \, \int_{\,\Omega_{\,a}} \ \theta_{\,n_{\,1}} \cdot \theta_{\,n_{\,2}} \ \mathrm{d}a \,.
\end{align*}
The matrix  relates the total sensitivity of the system. From its computation using the estimated parameters $\p^{\,\circ}$, a relative error estimator $\eta_{\,n} $ can be obtained for the parameter retrieved $p_{\,n}^{\,\circ}$ \citep{Walter_1982,Walter_1990}:
\begin{align*}
\eta_{\,n} \ \eqdef \ \frac{\sqrt{\bigl(\, F^{\,-1}\,\bigr)_{\,n\,n}}}{p^{\,\circ}_{\,n}} \,, \qquad \forall \, n \, \in \, \bigl\{\,1 \,,\, \ldots \,,\, N \,\bigr\}\,.
\end{align*}
High values of $\eta_{\,n}$ mean a high error during the estimation process.

\subsubsection{Solving the inverse problem combined with model selection}

The Approximate Bayesian Computation (ABC) algorithm is an efficient tool to infer the posterior distributions when the likelihood function is computationally too expensive to evaluate. It has been successfully implemented in various fields of research \citep{Toni_2009,Liepe_2014,daCosta_2018,Loiola_2020}. With the ABC technique, the prior information about the parameters is taken into account. In this way, it limits the variability of the estimated parameters in the inverse analysis.
Within the Bayesian framework, the objective is to approximate the posterior distribution $\prob{\p\,\bigl|\,\hat{u}}$ using \textsc{Bayes}' theorem:
\begin{align*}
 \prob{\p\,\bigl|\,\hat{u}} \egal \frac{\prob{\hat{u}\,\bigl|\,\p} \cdot \prob{\p}}{\prob{\hat{u}}} \,,
\end{align*}
where $\prob{\p}$ is the \emph{a priori} density of the parameters $\p^{\,\apr}\,$, $\prob{\hat{u}\,\bigl|\,\p}\,$ is the likelihood function and $\prob{\hat{u}}$ is the marginal probability density of measurements. 
The ABC algorithm can also be employed for model selection. An indicator $m$ is defined for each model of interest. Each model has a prior density $\prob{m}\,$. The marginal posterior distribution is approximated among all models and parameter subspaces, such as $\prob{m \,,\, \p\,\bigl|\,d\,\bigl(\,\hat{u}\,,\,u_{\,m}\,\bigr)\,\leqslant \,\varepsilon}\,$. So, it is useful to rank the models. Here, to increase the acceptance rate, the ABC method based on Sequential \textsc{Monte} \textsc{Carlo} sampling (SMC) is used. Before describing the details of the algorithm, several definitions are introduced.

First, a set of tolerances $\Omega_{\,\varepsilon}$ is specified by the user:
\begin{align*}
\Omega_{\,\varepsilon} \, \eqdef \, \Bigl\{\, 
\varepsilon_{\,i} \,\Bigr\} \,,
\quad \forall \ i \, \in \, \bigl\{\, 1 \,,\, \ldots \,,\, N_{\,\varepsilon}\,\bigr\}
\,,
\end{align*}
where $N_{\,\varepsilon}$ is the total number of tolerance populations sequentially generated by the method. The indicator $i$ corresponds to a population associated to tolerance $\varepsilon_{\,i}\,$. In this work, the distance function $d_{\,m}$ between the moisture content computed with the model $m$ and from experimental observation is defined as the square root of the sum of squared errors:
\begin{align}
\label{Eq:distance_function}
d_{\,m}\,\bigl(\,\p \,,\,  \hat{u} \,\bigr) \egal \biggl(\,
\int_{\,\Omega_{\,\hat{a}}} \, \Bigl(\, f_{\,m}\,\bigl(\,\p \,,\,\hat{a}\,\bigr) \moins \hat{u} \,\Bigr)^{\,2} \; \mathrm{d}a \,\biggr)^{\,\half}\,.
\end{align}
The nonzero systematic error has been omitted of the distance function due to its very low magnitude. Note that a discussion on the choice of the distance function is given in \cite[Section 2.1]{Toni_2009}. With Eq.~\eqref{Eq:distance_function}, a Maximum likelihood estimator is obtained. By assuming the measurement errors as \textsc{Gaussian} (as well as additive, unbiased and of constant variance), the estimator is unbiased, consistent, efficient and sufficient statistic \citep{Beck_1977}.

If the distance between a model $u_{\,m}$ and the experiments $\hat{u}$ is under tolerance $\varepsilon_{\,i}\,$, such model is a good candidate for being selected. Thus, we define the set of models $\Omega^{\,\circ}_{\,m}$ ``validating the distance test'' for the selected parameter:
\begin{align*}
\Omega^{\,\circ}_{\,m}  \, \eqdef \, \Bigl\{\, 
m \,\Bigl|\, d_{\,m}\,\bigl(\,\p \,,\,  \hat{u} \,\bigr) \, \leqslant \, \varepsilon_{\,i} \,, 
\ \forall \,  \varepsilon_{\,i} \, \in \, \Omega_{\,\varepsilon} \,,\, 
\,\Bigr\} \,.
\end{align*}
By analogy, the set of parameters $\Omega_{\,p}^{\,\circ}$ of models validating the distance test is:
\begin{align*}
\Omega_{\,p}^{\,\circ}  \, \eqdef \, \Bigl\{\, 
\p 
\,\Bigl|\, 
u_{\,m} \,,
\ \forall \, m \, \in \, \Omega_{\,m}^{\,\circ}
 \,\Bigr\} 
\,.
\end{align*}
The distance test is verified for a number of particles $N_{\,\nu}\,$. Each particle is identified by the indicator $j\,$.  The process is described in Algorithm~\ref{alg:ABC}. For each population $i$ and particle $j\,$, a model $m^{\,\star\,\star}$ is chosen according to its prior probability density. In this work, all the models are considered equally probable. Then, three main steps are highlighted. The first one handles the sampling of a candidate parameter $\p^{\,\star\,\star}\,$. It is performed according to a weight computed as:
\begin{align}
\label{eq:weights}
w_{\,i}^{\,j} \egal 
\begin{cases} 
\ 1 \,, & \quad j \egal 1 \,, \\
\ \displaystyle \frac{\prob{\p^{\,\star\,\star}}}
{\sum w_{\,i-1}^{\,j} \cdot \mathcal{K} \,\bigl(\,\p_{\,i-1}^{\,j} \,,\, \p^{\,\star\,\star} \,\bigr)} \,, & \quad j \ > \ 1 \,,
\end{cases} 
\end{align}
where $\p_{\,i-1}^{\,j}$ is the parameter obtained from previous population. Here, the perturbation kernel $\mathcal{K}$ is chosen as a random walk move with a uniform distribution $\mathcal{K} \,\bigl(\,\p\,,\,\p^{\,\star} \,\bigr) \egal \kappa \cdot \mathcal{U}\,\bigl(\, -1\, \,,\,1 \,\bigr)$ and $\displaystyle \kappa \egal \kappa_{\,0} \cdot \max_{\,\Omega_{\,p}} \p_{\,i}\,$ \citep{Toni_2009b}.
The second step carries the computation of the direct model for the water activity $\hat{a}\,$. The third step performs the distance test for the candidate model and its parameter. If it is successful, then the model $m^{\,\star\,\star}$ and parameter $\p^{\,\star\,\star}$ are stored in their respective subsets. In addition, the algorithm goes on for the next particle. If one of this three steps fails, it returns to Step~\ref{alg:step_sample_model} by again sampling the model. Before moving to the next population, the weights are normalized by performing the operation:
\begin{align}
\label{eq:norm_weights}
w_{\,i}^{\,j} \egal  \frac{w_{\,i}^{\,j}}{\displaystyle \sum_{j=1}^{N_{\,\nu}} \, w_{\,i}^{\,j}} \,.
\end{align}
Last, to evaluate the efficiency of the algorithm, the acceptance rate of the particle is denoted by $\tau\,$.

\begin{algorithm}
\caption{ABC SMC Algorithm.}\label{alg:ABC}
\begin{algorithmic}[1]
\State Set tolerances vector $\varepsilon$ and distance function $d$
\State Set population indicator $i \egal 1\,$
\State Set particle indicator $j \egal 1\,$
\While{$i\,\leqslant\,N_{\,\varepsilon}$}
\While{$j\,\leqslant\,N_{\,\nu}$}
\State Sample the model $m^{\,\star\,\star}  \, \in \, \Omega_{\,m}$ from $\prob{m}$ \label{alg:step_sample_model}
\If{$j \, \neq \, 1$} \Comment \emph{Sampling of a candidate parameter}
\State Sample $\p^{\,\star} \, \in \, \Omega_{\,p}^{\,\circ}$ with weight $w_{\,i-1}^{\,j}$
\State Perturb the parameter $\p^{,\star}$ to obtain $\p^{\,\star\,\star} \, \sim \, K \,\bigl(\,\p\,,\,\p^{\,\star} \,\bigr)$
\Else{ }
\State Sample $\p^{\,\star\,\star} \, \in \, \Omega_{\,p\,,\,m^{\,\star\,\star}}$ from $\prob{\p\,,\,m^{\,\star\,\star}}$
\EndIf 
\State \textbf{end}
\If{$\prob{\p^{\,\star\,\star}} \, \neq \, 0$} \Comment \emph{Computation of direct problem}
\State Compute $u_{\,m} \egal f_{\,m^{\,\star\,\star}}\;\bigl(\,\p^{\,\star\,\star} \,,\,\hat{a}\,\bigr) \,$
\Else{ }
\State Return to step~\ref{alg:step_sample_model}
\EndIf
\State \textbf{end}
\If{ $d_{\,m}\,\bigl(\,\p \,,\,  \hat{u} \,\bigr) \, \leqslant \, \varepsilon_{\,i}$ } \Comment \emph{Evaluating the distance test}
\State Store the candidate model $\Omega_{\,m^{\,\star\,\star}}^{\,\circ} \egal \Omega_{\,m^{\,\star\,\star}}^{\,\circ} \, \cup \, \{\, m^{\,\star\,\star} \,\}$
\State Store the associated parameters $\Omega_{\,p}^{\,\circ} \egal \Omega_{\,p}^{\,\circ} \, \cup \, \{\, p^{\,\star\,\star} \,\}$
\State Compute the weight $w_{\,i}^{\,j}$ using Eq.~\eqref{eq:weights}
\State $j \egal j + 1$
\Else{ }
\State Return to step~\ref{alg:step_sample_model}
\EndIf
\State \textbf{end}
\State Normalize the weight using Eq.~\eqref{eq:norm_weights}
\State $i \egal i + 1$ 
\EndWhile 
\State \textbf{end}
\EndWhile 
\State \textbf{end}
\end{algorithmic}
\end{algorithm}

\begin{figure}
\centering
\includegraphics[width=.95\textwidth]{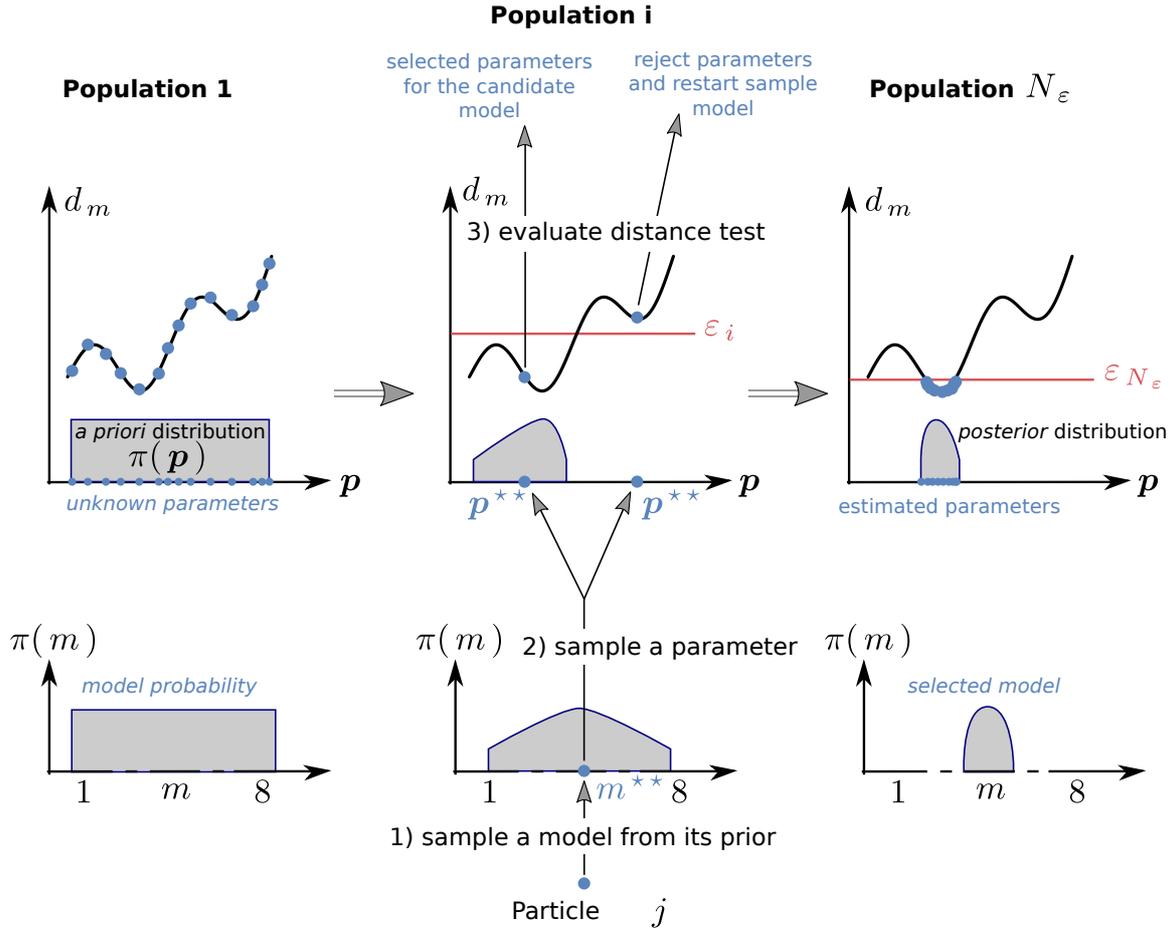}
\caption{Illustration of the principles of the ABC SMC algorithm.}
\label{fig:ABC}
\end{figure}

Figure~\ref{fig:ABC} summarizes the ABC algorithm used in this work. It proceeds as follows. At the beginning, the different models have uniform \emph{a priori} distributions, so that they have exactly the same probability of being selected. Uniform priors are also assigned to the parameters of each model. The algorithm runs for successive $N_{\,\varepsilon}$ populations with decreasing tolerances $\varepsilon\,$. For the population $i\,$, \emph{i.e.} the tolerance $\varepsilon_{\,i}\,$, the algorithm is divided in three main steps. \emph{(i)} First, sample a model $m^{\,\star\,\star}$ from the prior distribution among the eight competing models. If it is the first population, a uniform \emph{a priori} distributions are defined for the competing models and the unknown parameters. It enables to not favor any of the competing models. For the other populations, the priors for the parameters are obtained from the weights at the previous population. \emph{(ii)} Then, for the candidate model sampled from its prior, the algorithm samples a candidate parameter $p^{\,\star\,\star}\,$, using a sequential scheme and a small perturbation to ensure that the whole parameter space is explored. \emph{(iii)} For each candidate parameter, the distance $d_{\,m}$, \emph{i.e.} the error, between the direct model and the experiment is computed. Two cases are distinguished. If the distance is lower than the tolerance $\varepsilon_{\,i}\,$, then both the parameter and model are selected. It is stated that the particle validated the distance test. If the test is not valid, then the algorithm comes back to the first step in order to sample a new model from its prior. This operation is repeated until $N$ particles have been accepted, denoting by $\tau$ the acceptance rate. The tolerance of the last population corresponds to the desired agreement between the model and measurement data. Among the successive populations, less models validate the distance test since the tolerances are decreasing. At the end, only the best model(s) remain(s) and the samples for the parameters approximate their posterior distribution. In other words, the algorithm provides the best model and corresponding parameters values that minimize the error between measured and estimated quantities.

\section{Experiments}
\label{sec:experimental_measurement}

\subsection{Materials and methods}

The tested material is wood fibre insulation \citep{soprema} with density of $\rho \egal 50 \ \mathsf{kg\,.\,m^{\,-3}}$. Due to its low thermal conductivity and vapour permeability, this bio-based material is increasingly used in building envelope. Furthermore, it presents a strong hygroscopic behaviour as highlighted by the previous sorption isotherm measurement \citep{Vololonirina_2014}.

Sorption isotherms are measured with $5$ samples with the DVS equipment IGASorp-HT system (Hiden Isochema, Warrington, UK). The instrument has a microbalance with a resolution of $0.1 \ \mu\mathsf{g}$ on which a stainless-steel mesh basket containing the sample is suspended. The sample is then placed inside a separate chamber with controlled temperature and water activity, and sample mass is continuously recorded. Prior to the start of the adsorption measurement, sample is dried under flow of dry nitrogen at $65 \ \mathsf{^{\,\circ}C}$ for $24 \mathsf{h}$ (with a flow of $250 \mathsf{mL\,.\,min^{\,-1}}$) until stabilization \citep{Iso_12570}. Sample dry mass is recorded after setting the temperature to $23 \ \mathsf{^{\,\circ}C}$ under flow of dry nitrogen. Here, sample dry mass varies between $10.8$ and $28.2 \ \mathsf{mg}\,$. Then, the sample is exposed to increasing humidity from $0.05$ to $0.9$, with a $0.05$ step, the testing temperature being $23 \ \mathsf{^{\,\circ}C}\,$. The humidity is controlled by mixing dry and water vapour-saturated nitrogen streams at a total flow of $250 \mathsf{mL\,.\,min^{\,-1}}\,$ using electronic mass flow controllers. It is measured by a sensor placed in the chamber near the sample. The experiment is run at given temperature and water activity until an user defined stop criterion is reached. Usual stop criteria are \emph{(i)} hold time, \emph{(ii)} final rate of derivative of mass with respect to time $\displaystyle \od{m}{t}$ or \emph{(iii)} accuracy of asymptotic moisture content from a kinetic model fit to the moisture content versus time data. While several stop criteria may be found in the literature for cellulosic materials, Glass et al. \citep{Glass_2017} underlined that the commonly used stop criteria may mischaracterize equilibrium moisture content up to $1 \ \mathsf{\%}$ of the moisture content. Therefore, the recommendation is to increase hold times to catch long time constants of sorption kinetic (the order of $500 \ \mathsf{min}$ or longer), even if it increases the isotherm measurement time to several weeks for a single replicate of a single material. Later, the same authors \citep{Glass_2018} proposed a new methodology to improve measurement accuracy and to reduce measurement times. Based on these previous works, a slope $\displaystyle \od{m}{t} \egal 10 \ \mathsf{\mu g\,.\,g^{\,-1}\,.\,min^{\,-1}}$ calculated over a $15 \ \mathsf{min}$ window combined with a maximum hold time of $24 \ \mathsf{h}$ was used as stop criterion. When this condition is met, the apparent equilibrium moisture content is taken as the last measured moisture content.

\subsection{Experimental data}

\begin{figure}
\centering
\subfigure[\label{fig:S_fa_all}]{\includegraphics[width=.45\textwidth]{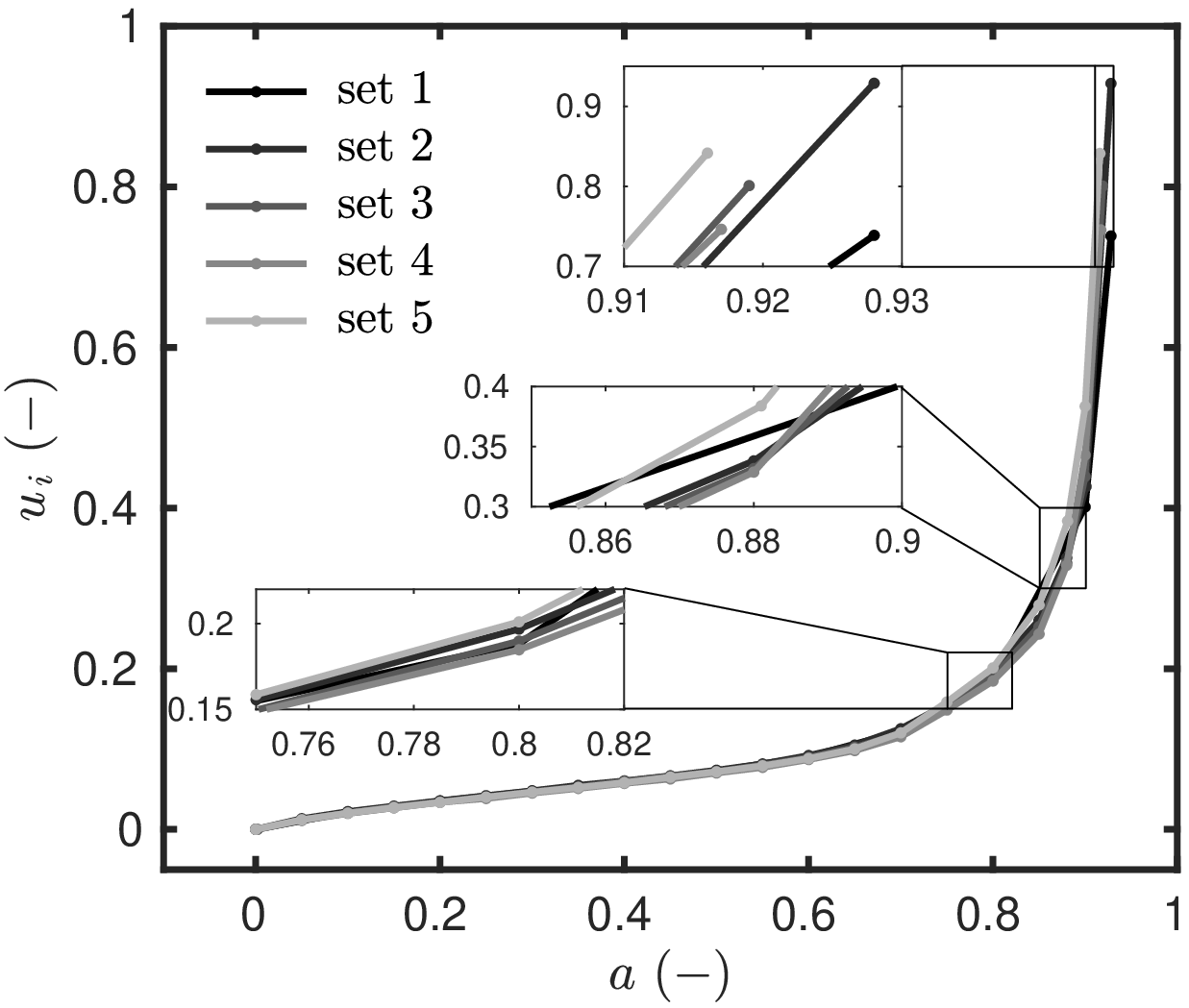}}  \hspace{0.2cm}
\subfigure[\label{fig:Sigma_fa}]{\includegraphics[width=.45\textwidth]{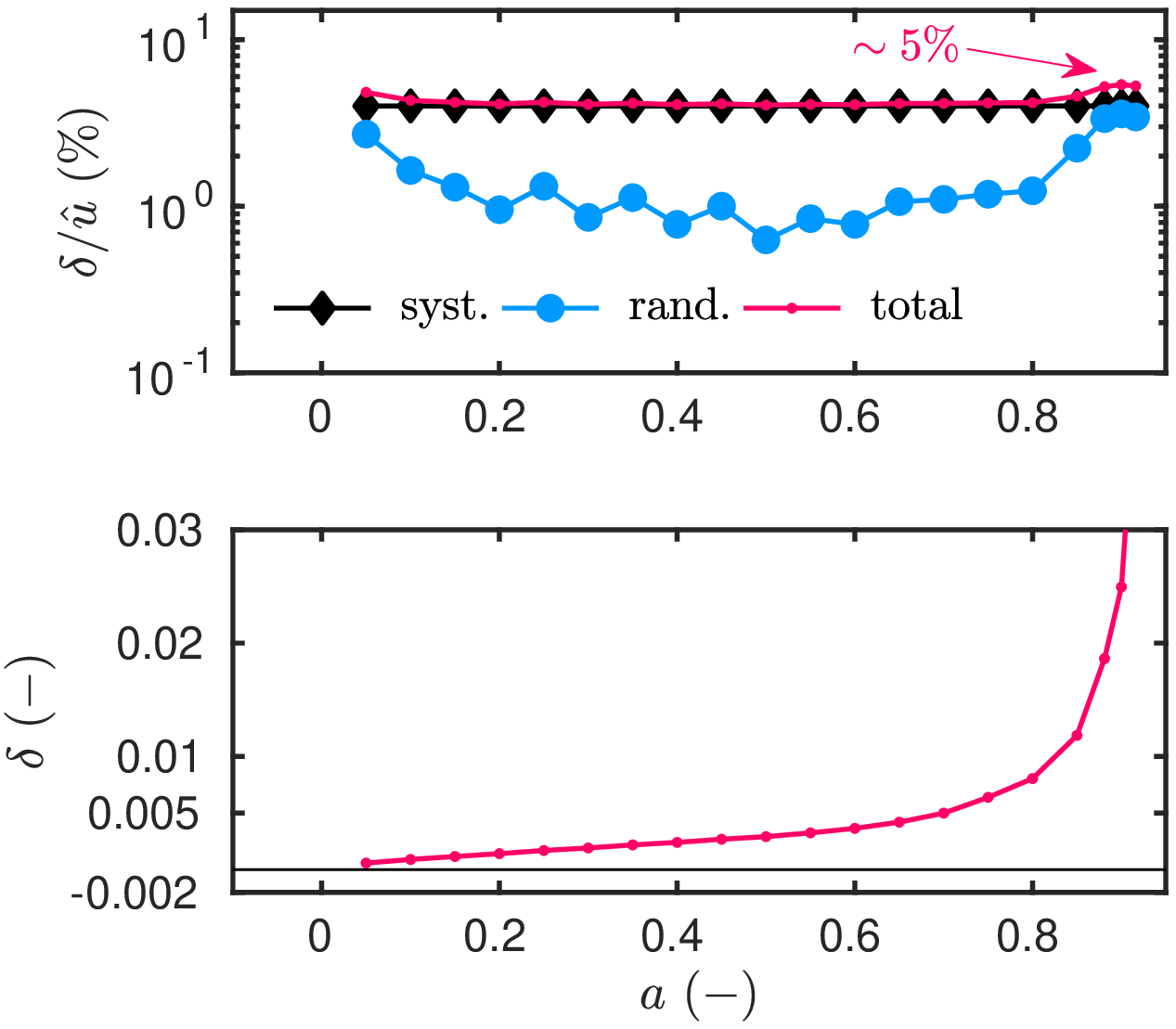}} \\
\subfigure[\label{fig:Sbest_fa}]{\includegraphics[width=.93\textwidth]{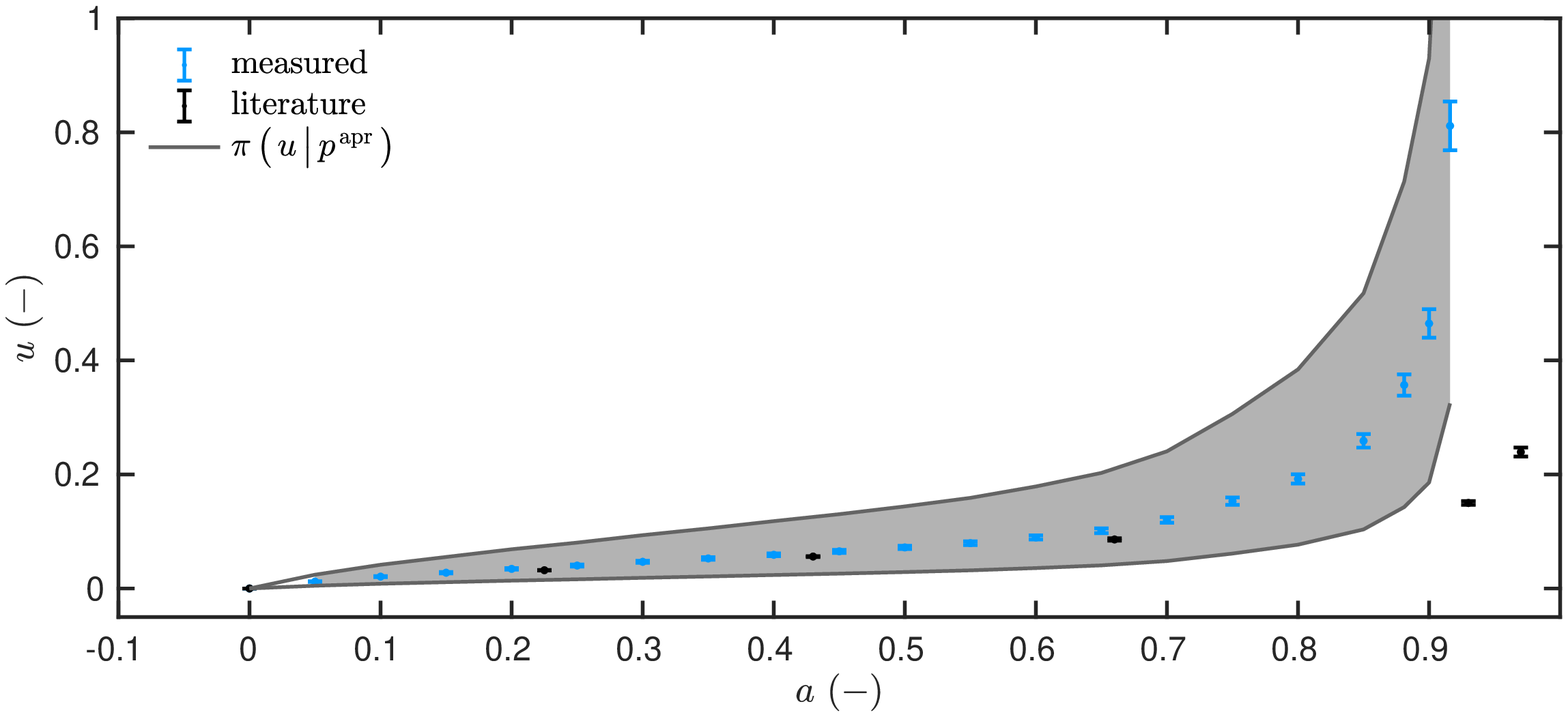}}
\caption{Measured moisture content in the wood fibre material for the five samples \emph{(a)}. Variation of the measurement uncertainty according to the water activity \emph{(b)}. Best estimate for the measured moisture content compared to results from the literature \citep{Vololonirina_2014} \emph{(c)}.}
\label{fig:S_fa}
\end{figure}

The experimental results for the five samples are presented in Figure~\ref{fig:S_fa_all}. Except for three points with water activity higher than $0.9$, all equilibrium moisture contents are obtained by meeting the stop criterion. The measurement time increases with water activity, ranging between $2$ and $12 \ \mathsf{h}\,$. In the hygroscopic range (\emph{i.e.} $a \,<\, 0.8$), acquisition could be stopped because of high signal-to-noise ratio. Nevertheless, the calculated slope over a $60 \ \mathsf{min}$ window did not exceed $15 \ \mathsf{\mu g\,.\,g^{\,-1}\,.\,min^{\,-1}}\,$. Therefore, we have good confidence in the results in the hygroscopic range and the discrepancy is limited. For water activity higher than 0.8, the signal to noise ratio is better because of the larger mass change. Nevertheless, even if each measurement lasts for at least $4 \mathsf{h}$, it might not be sufficient for the identification of long time constants of sorption kinetic. For instance, the calculated slope over a $120 \ \mathsf{min}$ window did not drop below $45 \ \mathsf{\mu g\,.\,g^{\,-1}\,.\,min^{\,-1}}\,$, which is much higher than the
value of $3 \ \mathsf{\mu g\,.\,g^{\,-1}\,.\,min^{\,-1}}$ suggested by \textsc{Glass} et al. \citep{Glass_2018}. Therefore, we expect measuring moisture content with an accuracy of at least $0.36 \ \%$ \citep{Glass_2018}. Nevertheless, the sorption at high water activity involves a complex phenomenon (like polymer softening) that may vary from one sample to another. This variability is highlighted by the higher discrepancy between the sets for water activity higher than $0.8\,$.

According to \textsc{Taylor} \citep{Taylor_1997}, the best estimates of the moisture content is:
\begin{align*}
\hat{u} \ \eqdef \ \frac{1}{N_{\,e}} \, \sum_{i \egal 1}^{N_{\,e}} \, u_{\,i} \,,
\end{align*}
where $N_{\,e}$ is the number of carried out measurements. In the present case, $N_{\,e} \egal 5$ since measurement for five samples has been taken. To evaluate the total measurement uncertainty $\delta$, both the random and the systematic components of the uncertainty are considered:
\begin{align*}
\delta \ \eqdef \ \sqrt{\delta_{\,\sim}^{\,2} \plus \delta_{\,\Sigma}^{\,2} } \,.
\end{align*}
The random part $\delta_{\,\sim}$ is computed through the standard deviation of the mean:
\begin{align*}
\delta_{\,\sim} \ \eqdef \ \frac{1}{\sqrt{N_{\,e}}} \ 
\sqrt{ 
\frac{1}{N_{\,e}} \, \sum_{i \egal 1}^{N_{\,e}} \, \bigl(\, u_{\,i} \moins \hat{u} \,\bigr)^{\,2} \,.
 }
\end{align*}
The systematic component $\delta_{\,\Sigma}$ is due to the experimental DVS device and given as follows:
\begin{align*}
\delta_{\,\Sigma} \egal 10^{\,-7} \,.
\end{align*}
Here, it is given by the balance resolution divided by the dry mass. It is assumed that there is no other systematic error in the measurement design.
The best estimate for the moisture content is given in Figure~\ref{fig:Sbest_fa} as well as in Table~\ref{tab:S_fa}. The different uncertainty components are also indicated. The variation of the relative uncertainty with the water activity is shown in Figure~\ref{fig:Sigma_fa}. It can be noticed that the random uncertainty component is significant compared to the systemic one. In Figure~\ref{fig:Sbest_fa}, it seems that the uncertainty increases at high water activity. However, from a relative point of view, the uncertainty is high for $a \, \geqslant \, 0.8\,$, of the order of $5 \% \,$. As samples have different dry masses and shapes, it may explain the observed higher uncertainty.
In Figure~\ref{fig:Sbest_fa}, the measured sorption isotherm is compared to previous results \citep{Vololonirina_2014} from the literature. The shape in the hygroscopic domain is similar. For higher water activity, large difference may be observed, probably due to differences in stop criteria. Indeed, if the measurement is stopped too early, it may lead to an underestimation of equilibrium moisture content during adsorption \citep{Glass_2017}. 

\begin{table}
\centering
\caption{Best estimate for the moisture content according to the water activity in the wood fibre material.}
\label{tab:S_fa}
\setlength{\extrarowheight}{.5em}
\begin{tabular}[l]{@{} cc cc c}
\hline
\hline
\textit{Water activity}
& \textit{Moisture content} 
& \textit{Total uncertainty} 
& \textit{Random uncertainty}  \\
$a \ \unit{-}$ 
& $\hat{u}\ \unit{-}$
& $\delta\ \unit{-}$
& $\delta_{\,\sim}\ \unit{-}$ \\
\hline 
0 & 0 & 0.01 & 0 \\
0.05 & 0.0121 & 0.0006 & 0.0005 \\
0.1 & 0.0208 & 0.0009 & 0.0008 \\
0.15 & 0.0276 & 0.0012 & 0.0011 \\
0.2 & 0.0344 & 0.0014 & 0.0014 \\
0.25 & 0.0402 & 0.0017 & 0.0016 \\
0.3 & 0.0468 & 0.0019 & 0.0019 \\
0.35 & 0.0527 & 0.0022 & 0.0021 \\
0.4 & 0.059 & 0.0024 & 0.0024 \\
0.45 & 0.0652 & 0.0027 & 0.0026 \\
0.5 & 0.072 & 0.0029 & 0.0029 \\
0.55 & 0.0795 & 0.0032 & 0.0032 \\
0.6 & 0.0895 & 0.0036 & 0.0036 \\
0.65 & 0.1013 & 0.0042 & 0.0041 \\
0.7 & 0.1203 & 0.005 & 0.0048 \\
0.75 & 0.1531 & 0.0064 & 0.0061 \\
0.8 & 0.1921 & 0.008 & 0.0077 \\
0.85 & 0.2589 & 0.0119 & 0.0143 \\
0.881 & 0.3568 & 0.0186 & 0.012 \\
0.9 & 0.4647 & 0.025 & 0.0186 \\
0.916 & 0.8113 & 0.0428 & 0.0325 \\
\hline
\hline
\end{tabular}
\end{table}

\subsection{A priori distribution of unknown parameters}

The slope of the sorption model MADS in Eq.~\eqref{eq:MADS_model} is $K \egal 0.2416\,$. Uniform distributions are considered for the \emph{prior} density of the unknown parameters. The interval of variation of each parameter is given in Table~\ref{tab:prior_distribution_parameters} for each model. To challenge each of the competing models, the interval of variation of the \emph{prior} density is defined to represent a large range of sorption curves.  The interval of variation of the \emph{a priori} parameters are chosen so that the image of each model is included in the range of sorption curves illustrated in Figure~\ref{fig:Sbest_fa}. It should be noted that the magnitude of variation of the parameters is very different among the models. It will be verified that those differences do not influence the model selection through the choice of the kernel parameter.

\begin{table}
\centering
\caption{\emph{Prior} distribution of the unknown parameters for each model.}
\label{tab:prior_distribution_parameters}
\setlength{\extrarowheight}{.5em}
\begin{tabular}[l]{@{} c cccc}
\hline
\hline
\textit{Models}
& \textit{Parameter} $p_{\,1}$
& \textit{Parameter} $p_{\,2}$
& \textit{Parameter} $p_{\,3}$ 
& \textit{Parameter} $p_{\,4}$ \\
\hline 
GAB
& $\mathcal{U}\,\bigl(\, 1.06 \e{-2} \,,\, 5.31 \e{-2} \,\bigr)$
& $\mathcal{U}\,\bigl(\, 0.95 \,,\, 1.05 \,\bigr)$
& $\mathcal{U}\,\bigl(\, 5 \,,\, 171 \,\bigr)$
& -\\
TRM
& $\mathcal{U}\,\bigl(\, 1.0 \,,\, 3.0\,\bigr)$
& $\mathcal{U}\,\bigl(\, 1.0 \,,\, 1.55\,\bigr)$
& $\mathcal{U}\,\bigl(\, 2.0 \,,\, 3.0\,\bigr)$
& -\\
OSW
& $\mathcal{U}\,\bigl(\, 0.013 \,,\, 0.14\,\bigr)$
& $\mathcal{U}\,\bigl(\, 0.75 \,,\, 1.28\,\bigr)$
& - \\
FX
& $\mathcal{U}\,\bigl(\, 4.9 \,,\, 40\,\bigr)$
& $\mathcal{U}\,\bigl(\, 18.0 \,,\, 18.69\,\bigr)$
& $\mathcal{U}\,\bigl(\, 7 \,,\, 18.5\,\bigr)$ 
& $\mathcal{U}\,\bigl(\, 1.46 \,,\, 1.7\,\bigr)$ \\
BET
& $\mathcal{U}\,\bigl(\, 0.034 \,,\, 0.09 \,\bigr)$
& $\mathcal{U}\,\bigl(\, 0.20\,,\, 10\,\bigr)$
& -& -\\
VG
& $\mathcal{U}\,\bigl(\, 0.99 \,,\, 179 \,\bigr)$
& $\mathcal{U}\,\bigl(\, 26.6 \,,\, 3.73 \e{3}\,\bigr)$
& $\mathcal{U}\,\bigl(\, 1.7 \,,\, 2.4\,\bigr)$
& -\\
SM
& $\mathcal{U}\,\bigl(\, 0.0026 \,,\, 0.013 \,\bigr)$
& $\mathcal{U}\,\bigl(\, 0.04 \,,\, 0.20\,\bigr)$
& -
& -\\
MADS
& $\mathcal{U}\,\bigl(\, -1.1 \,,\, -0.4 \,\bigr)$
& $\mathcal{U}\,\bigl(\, 2.1 \,,\, 2.9\,\bigr)$
& -
& -\\
\hline
\hline
\end{tabular}
\end{table}

\section{Reliability of the models}
\label{sec:reliability_models}

The purpose is to demonstrate that the unknown parameters of the sorption models are identifiable.
From a theoretical point of view, the Structural Global Identifiability (SGI) property is evaluated for each model in Appendix~\ref{app:structural_identifiability} in Electronic Supplementary Material. As a synthesis, it is demonstrated that all eight models have parameters theoretically identifiable if a set of observations is obtained. The next section investigates the primary identifiability, \emph{i.e.} if the parameters sufficiently influence the output of the model to be estimated with accuracy.

\subsection{Primary identifiability}

\begin{figure}
\centering
\subfigure[\label{fig:TmGAB_fa} $p \egal \bigl(\,1.06 \e{-2}\,,\,0.95\,,\,5\,\bigr) $]{\includegraphics[width=.45\textwidth]{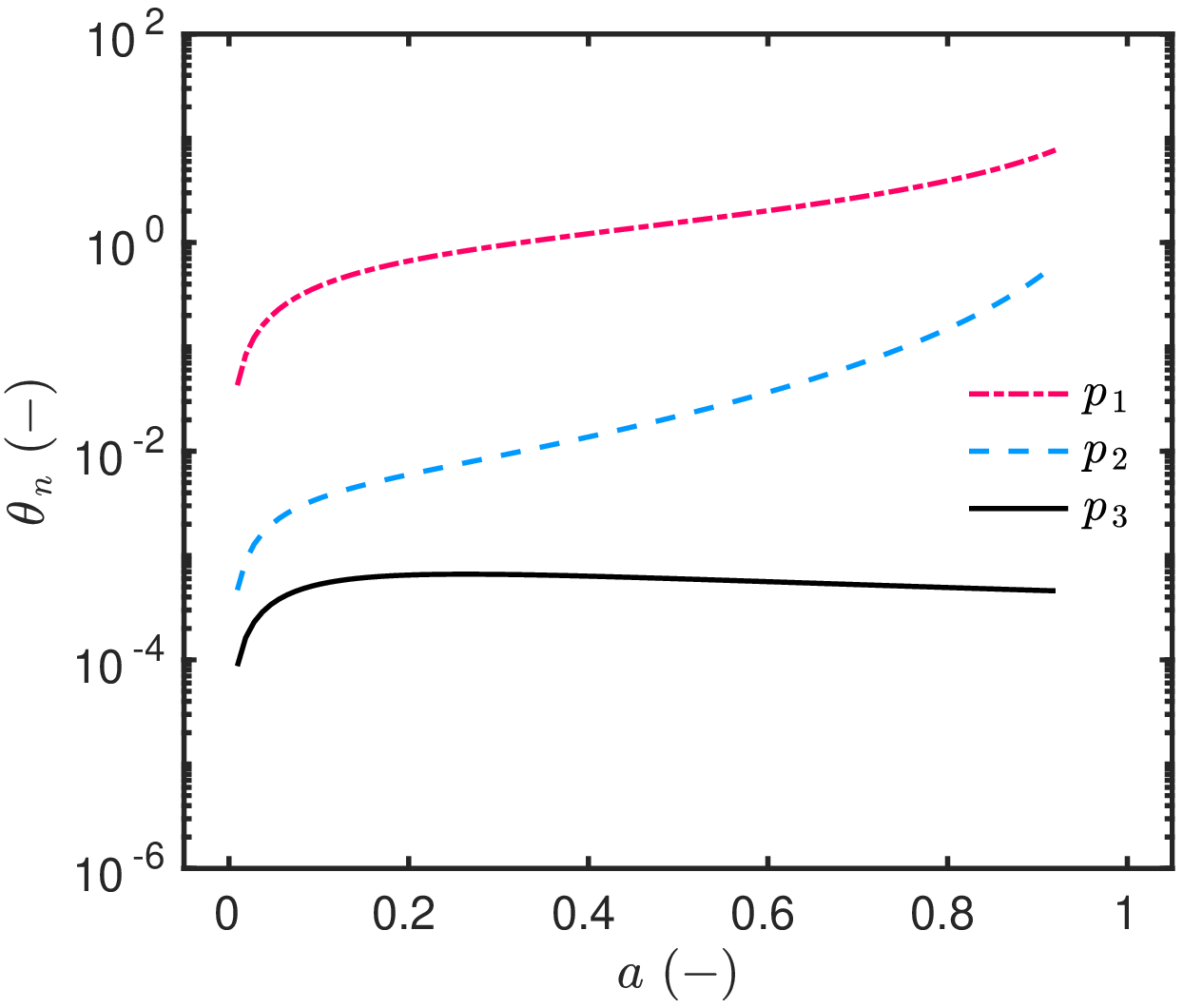}} \hspace{0.2cm}
\subfigure[\label{fig:TpGAB_fa} $p \egal \bigl(\,5.31 \e{-2}\,,\,1.05 \,,\,171\,\bigr)$]{\includegraphics[width=.45\textwidth]{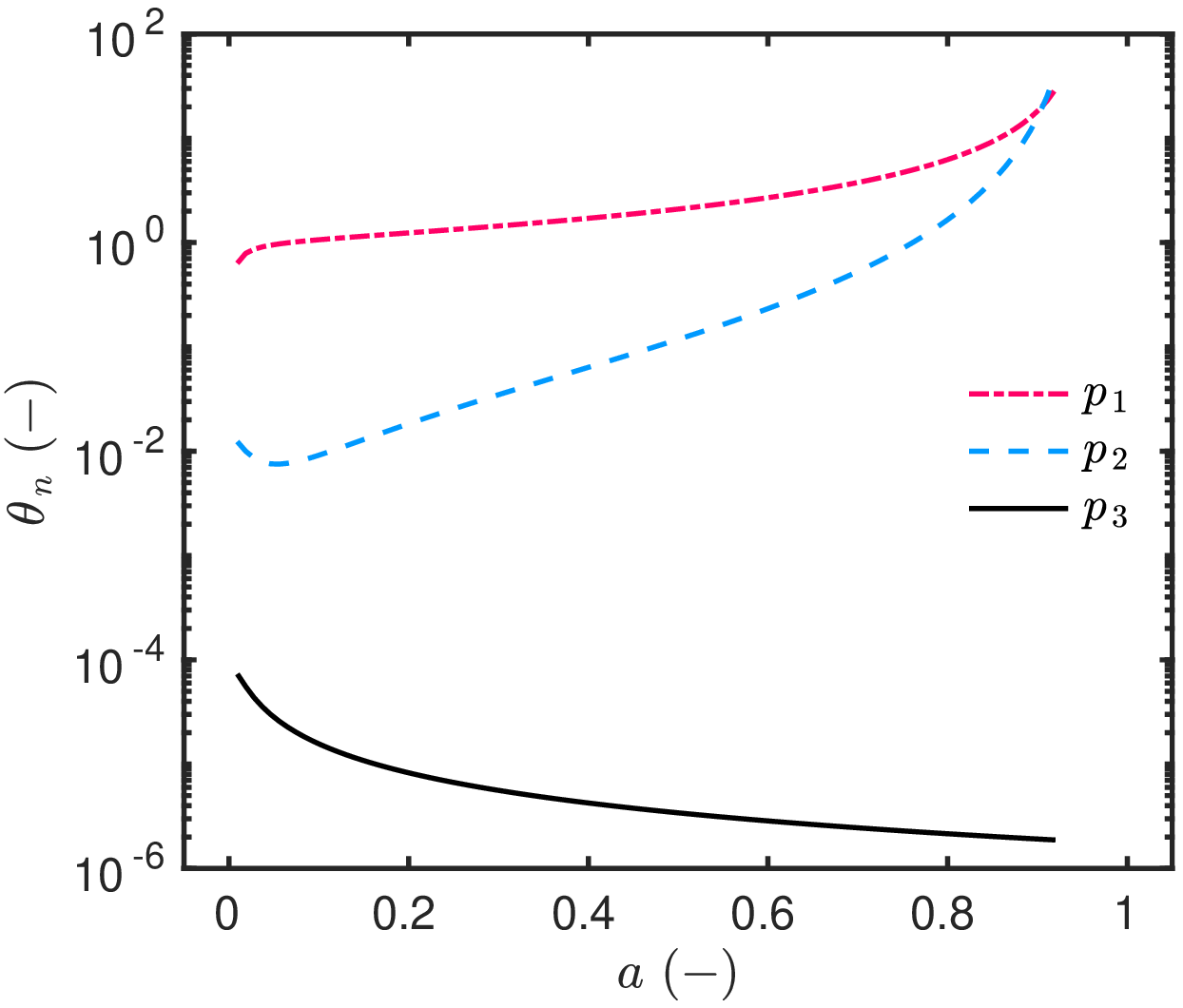}} 
\caption{Variation of the sensitivity functions of the GAB \emph{(a,b)} model.}
\label{fig:theta}
\end{figure}

\begin{figure}
\centering
\subfigure[\label{fig:gamma_GAB} GAB]{\includegraphics[width=.45\textwidth]{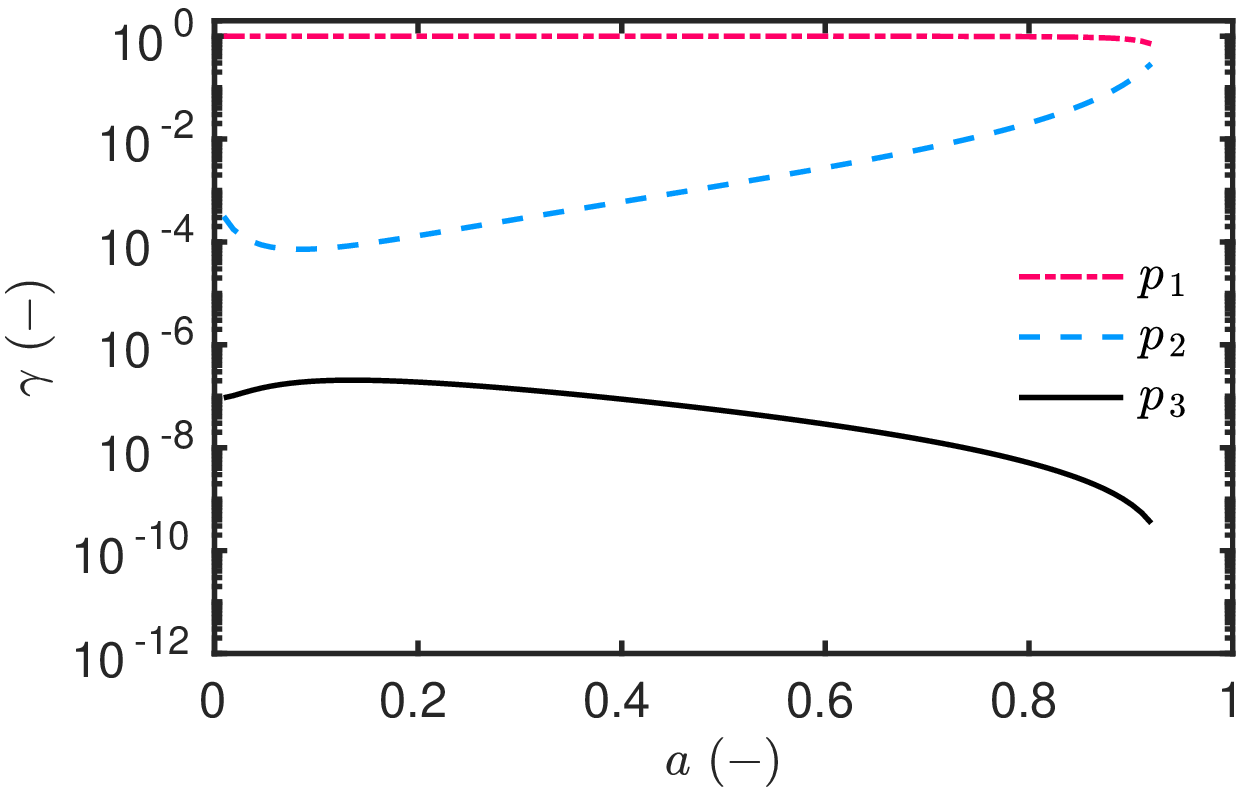}}  \hspace{0.2cm}
\subfigure[\label{fig:gamma_TRM} TRM]{\includegraphics[width=.45\textwidth]{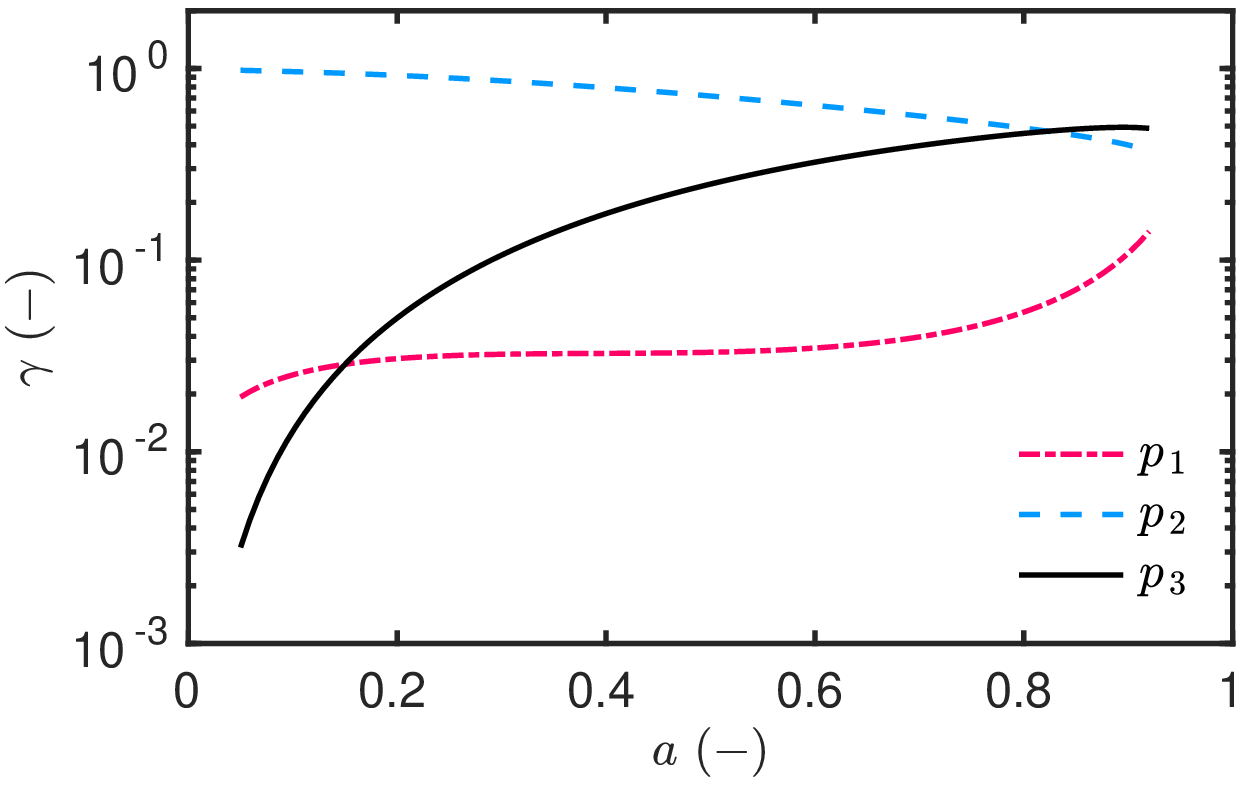}} \\
\subfigure[\label{fig:gamma_OSW} OSW]{\includegraphics[width=.45\textwidth]{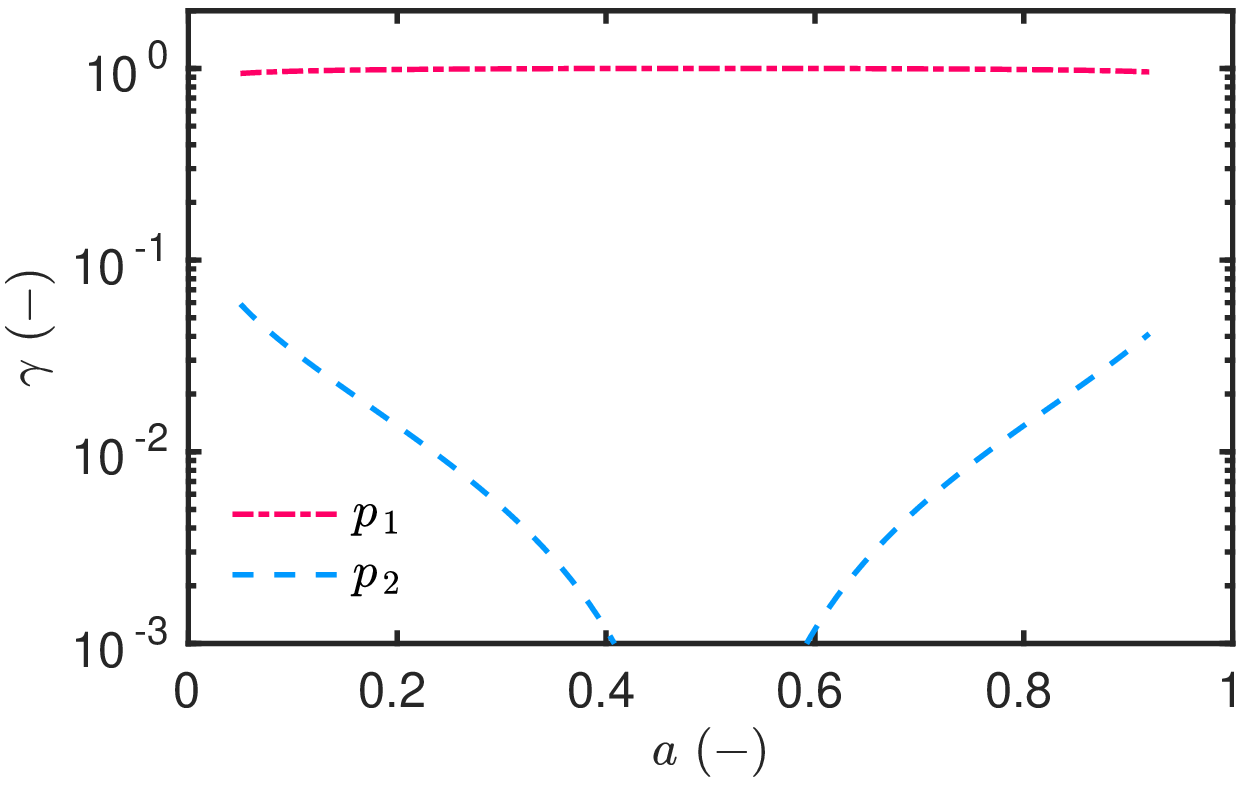}} \hspace{0.2cm}
\subfigure[\label{fig:gamma_FX} FX]{\includegraphics[width=.45\textwidth]{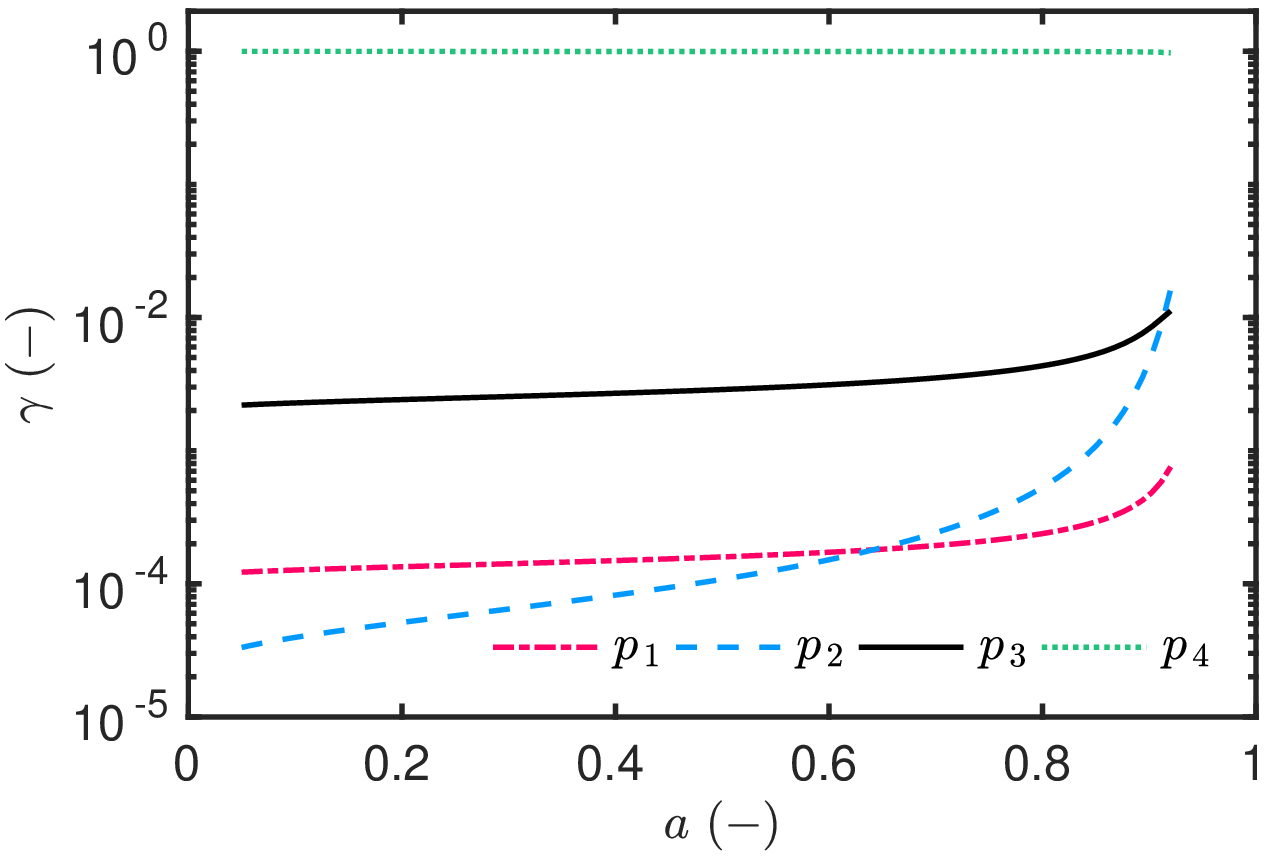}} \\
\subfigure[\label{fig:gamma_BET} BET]{\includegraphics[width=.45\textwidth]{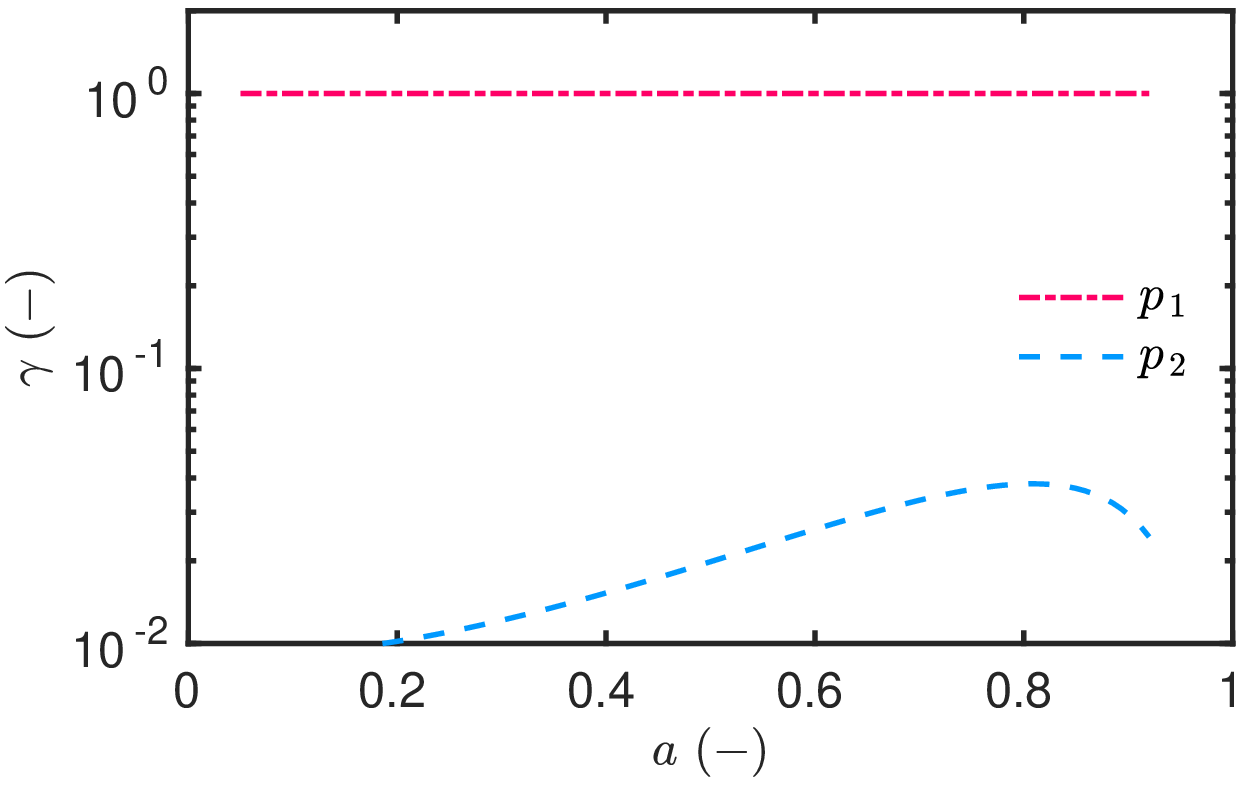}} \hspace{0.2cm}
\subfigure[\label{fig:gamma_VG} VG]{\includegraphics[width=.45\textwidth]{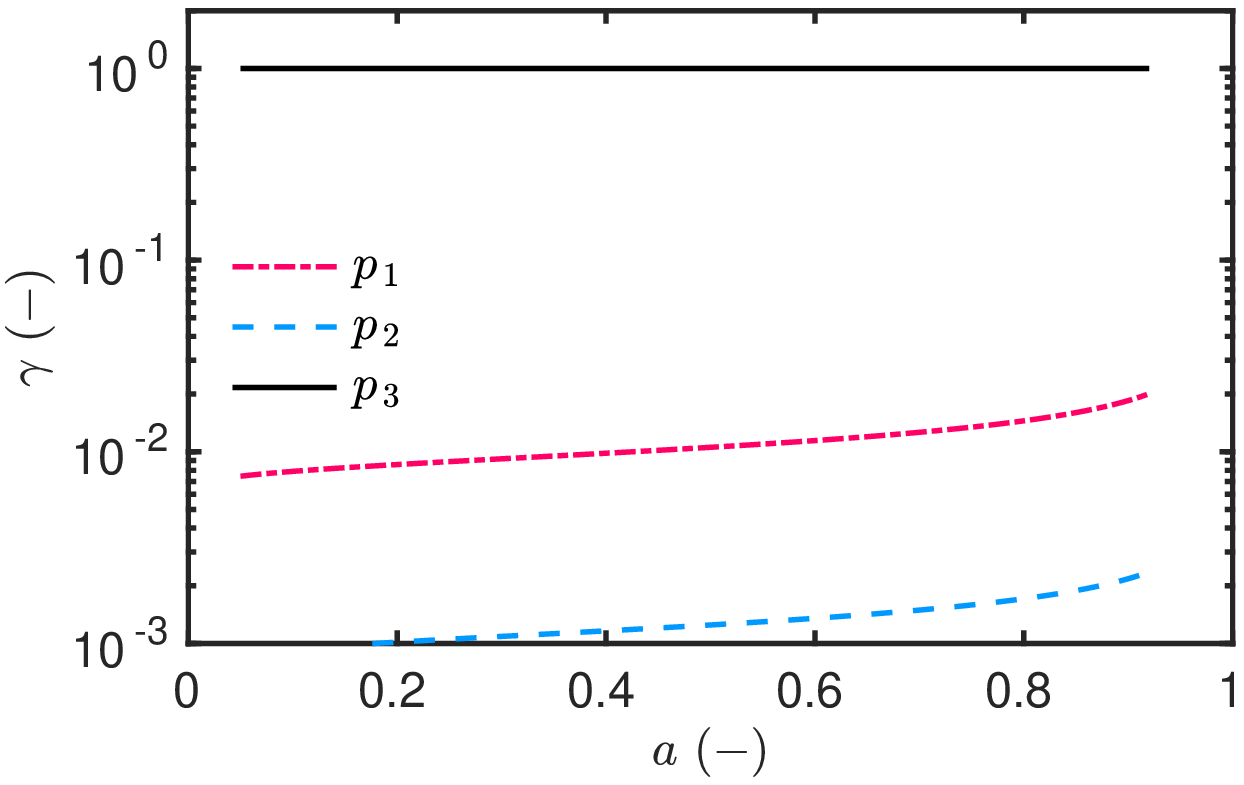}} \\
\subfigure[\label{fig:gamma_SM} SM]{\includegraphics[width=.45\textwidth]{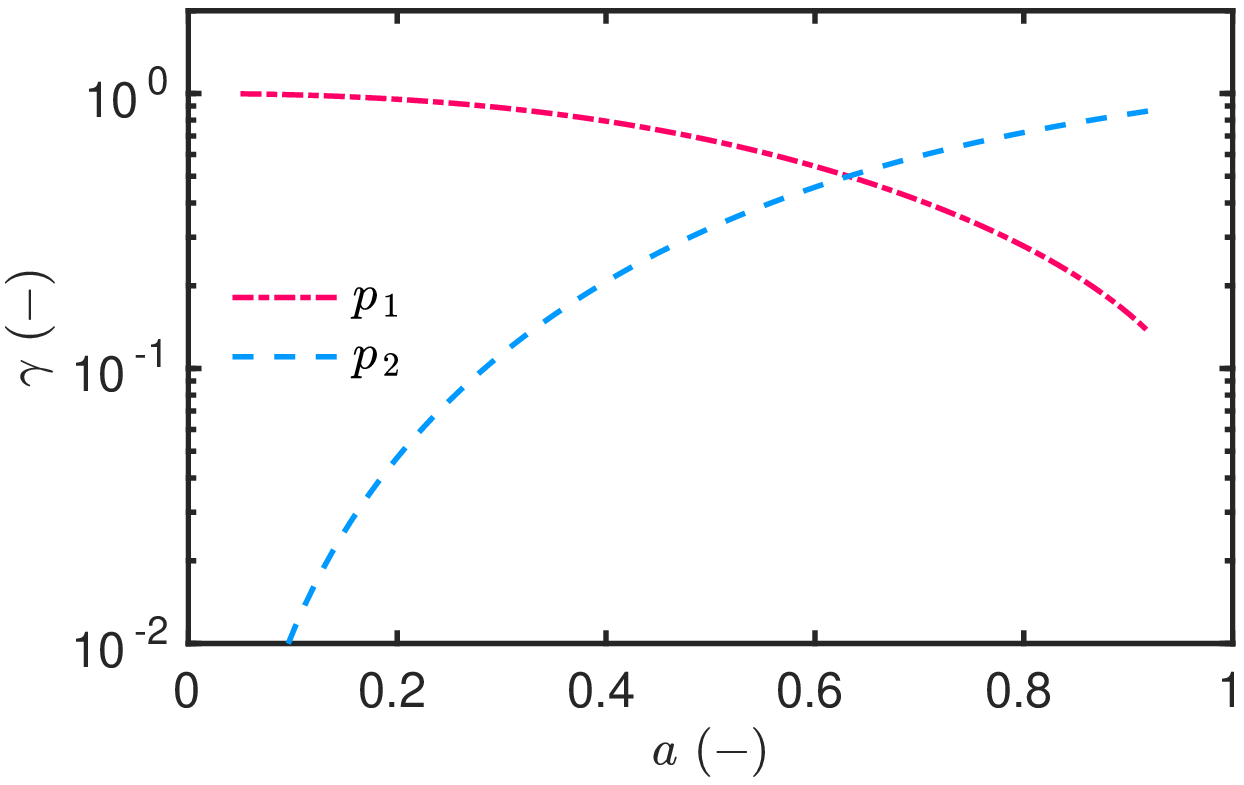}} \hspace{0.2cm}
\subfigure[\label{fig:gamma_MADS} MADS]{\includegraphics[width=.45\textwidth]{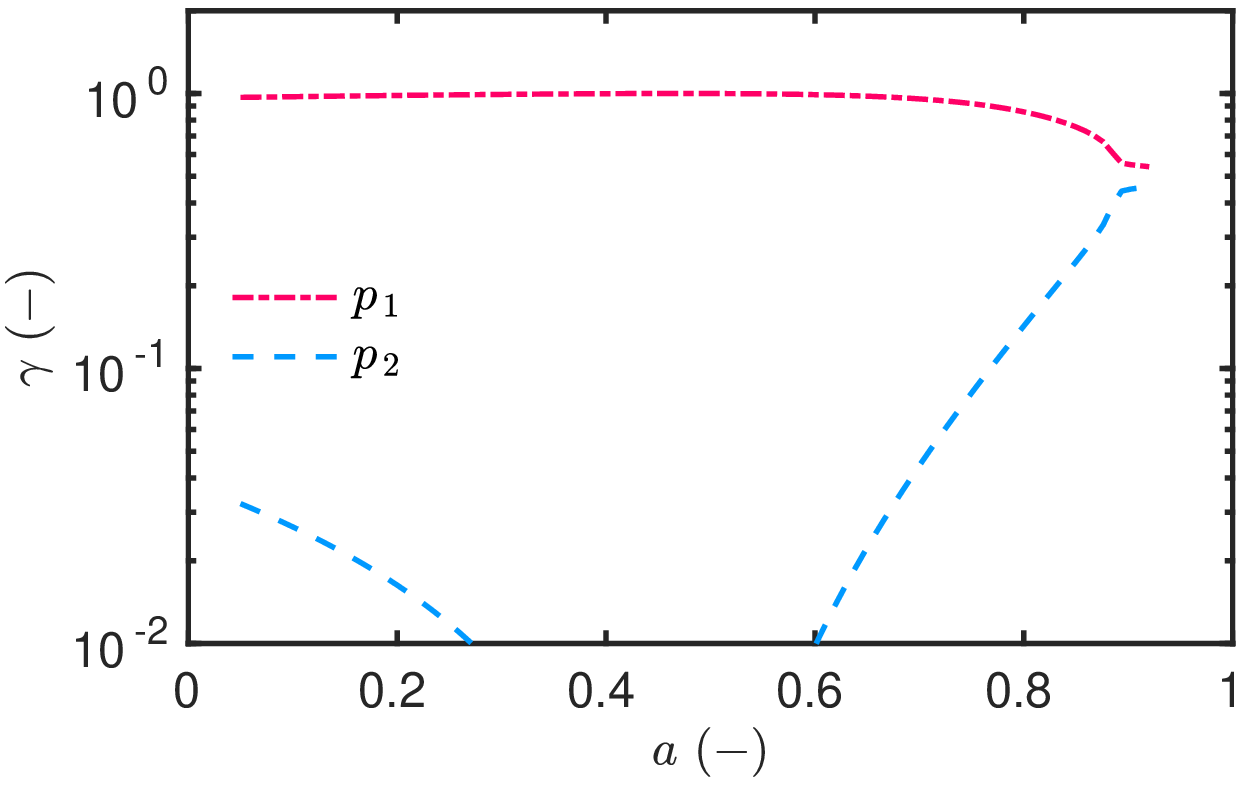}}
\caption{Variation of the global sensitivity metrics for the models according to the water activity.}
\label{fig:gamma}
\end{figure}

The sensitivity function is computed for the eight models. An illustration is shown for the GAB model in Figures~\ref{fig:TmGAB_fa} and \ref{fig:TpGAB_fa}, for two values of parameter $p\,$. The latter corresponds to the lower and upper bounds of each parameter in $\Omega_{\,p}$ for the GAB model, according to Table~\ref{tab:prior_distribution_parameters}. It is noticed that over the domain $\Omega_{\,a}\,$, the sensitivity functions of parameters $p_{\,1}$ and $p_{\,2}$ have the highest values. The sensitivity function of the parameter $p_{\,3}$ is at least $2$ orders of magnitude lower than others. These results are consistent with the sensitivity metric $\gamma$ computed for each of the three parameters and presented in Figure~\ref{fig:gamma_GAB}. The parameter $p_{\,3}$ has a very negligible influence on the sorption model. The sensitivity of parameters $p_{\,1}$ and $p_{\,2}$ is higher, with $\gamma_{\,1} \, > \, \gamma_{\,2}\,$. Thus, the parameter $p_{\,1}$ is the most sensitive of the model. Therefore, it is the easiest to identify from a practical point of view. As the water activity $a$ increases, the sensitivity of the parameter $p_{\,2}$ increases. Thus, one could imagine to use some observations for $a \, \in \, \bigl[\,0.05 \,,\, 0.8 \,\bigr]$ to estimate the parameter $p_{\,1}$ and then, some observations for $a \, \in \, \bigl[\,0.8 \,,\, 0.95 \,\bigr]$ to retrieve $p_{\,2}\,$. Even with this procedure, the parameter $p_{\,3}$ cannot be estimated with accuracy. 

The variation of the sensitivity metric $\gamma$  with $a$ is given for the other models in Figures~\ref{fig:gamma_TRM} to \ref{fig:gamma_VG}. The results of the primary identifiability are reported in Table~\ref{tab:primary_identifiability}. A general observation is that $5$ models out of $8$ have one parameter with a very high sensitivity, other parameters having a very low influence on the models output. In other words, these models have parameters with sensitivity metrics of the same orders of magnitude. One can conclude that their primary identifiability is very low for all the domain of $a\,$. The accuracy of the results of the parameter estimation problem might be very low. Three models, namely SM, TRM and MADS, have parameters with medium sensitivity for two parameters. For the SM and MADS model, both parameters have a similar influence on the model output. It shows a good primary identifiability for these models. For the TRM, parameters $p_{\,2}$ and $p_{\,3}$ have good primary identifiability. However, the accuracy of estimation for parameter $p_{\,1}$ might be poor since it has a very small influence on the model.

\begin{table}
\centering
\caption{Results of the primary identifiability for each model.}
\label{tab:primary_identifiability}
\setlength{\extrarowheight}{.5em}
\begin{tabular}[l]{@{} c cccc}
\hline
\hline
\multirow{2}{*}{\textit{Models}}
& \multicolumn{4}{c}{Global sensitivity indexes $\gamma^{\,\intercal}$} \\
& $p_{\,1}$
& $p_{\,2}$
& $p_{\,3}$
& $p_{\,4}$ \\
\hline 
GAB
& $0.89$
& $0.10$
& $\mathcal{O}(\,10^{\,-6}\,)$
& -\\
TRM
& $0.06$
& $0.55$
& $0.38$
& -\\
OSW
& $0.975$
& $0.025$
& - & - \\
FX
& $3\e{-4}$
& $3.7\e{-3}$
& $6.2\e{-3}$
& $0.98$ \\
BET
& $0.99$
& $3.2\e{-4}$
& -& -\\
VG
& $7.7\e{-2}$
& $9\e{-4}$
& $0.99$
 & -\\
SM
& $0.48$
& $0.52$
& -
 & -\\
MADS
& $0.54$
& $0.46$
& -
 & -\\
\hline
\hline
\end{tabular}
\end{table}

\subsection{Parameter estimation and model selection}
\label{sec:parameter_estimation_problem}

Previous investigations demonstrated that all models have parameters identifiable in theory. From a practical point of view, it has been shown that not all parameters of the models have sufficient influence on the model predictions to be retrieved with accuracy. Here, the parameter estimation problem is first solved to examine the consequence of a bad primary identifiability on the model reliability. Then, a selection is operated over the competing models to distinguish the most reliable ones.

\subsubsection{Parameter estimation}

\begin{figure}
\centering
\subfigure[\label{fig:S_fa_GAB_BET}]{\includegraphics[width=.45\textwidth]{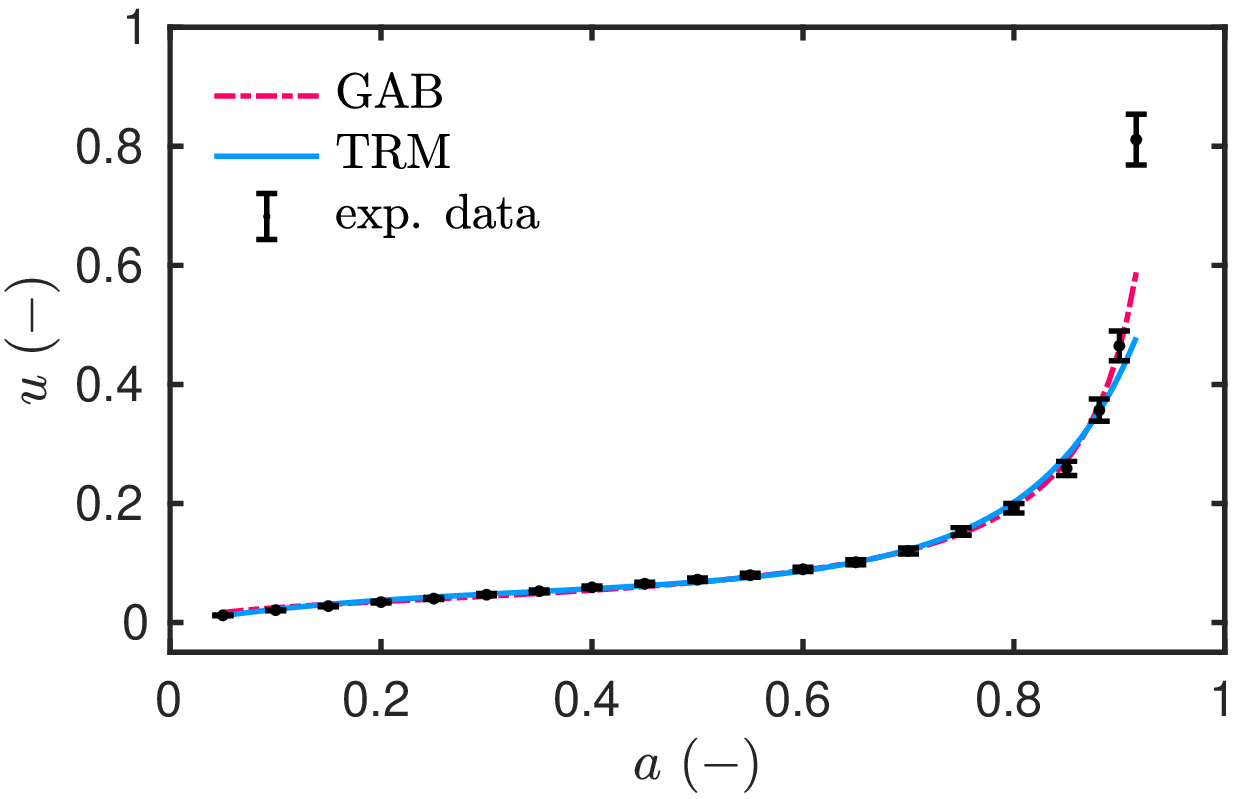}}  \hspace{0.2cm}
\subfigure[\label{fig:res_fa_GAB_BET}]{\includegraphics[width=.45\textwidth]{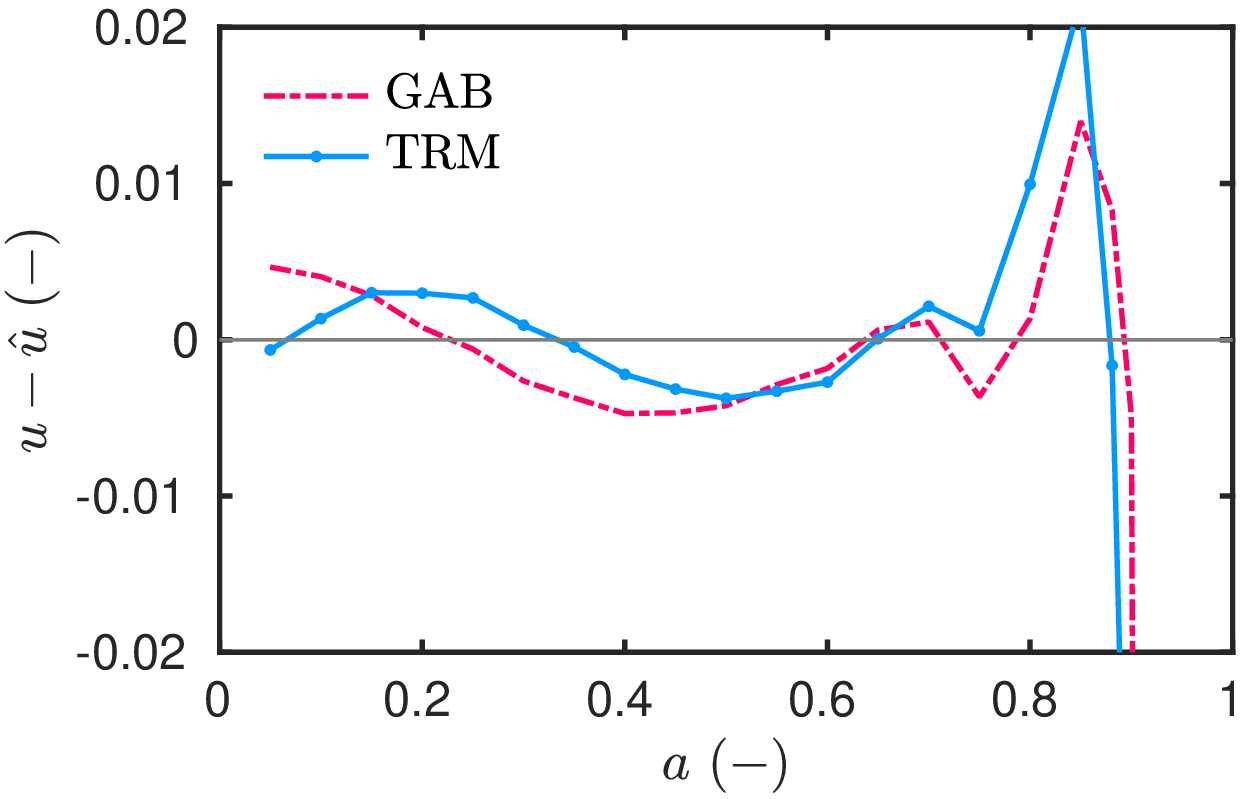}} \\
\subfigure[\label{fig:S_fa_OSW_FX}]{\includegraphics[width=.45\textwidth]{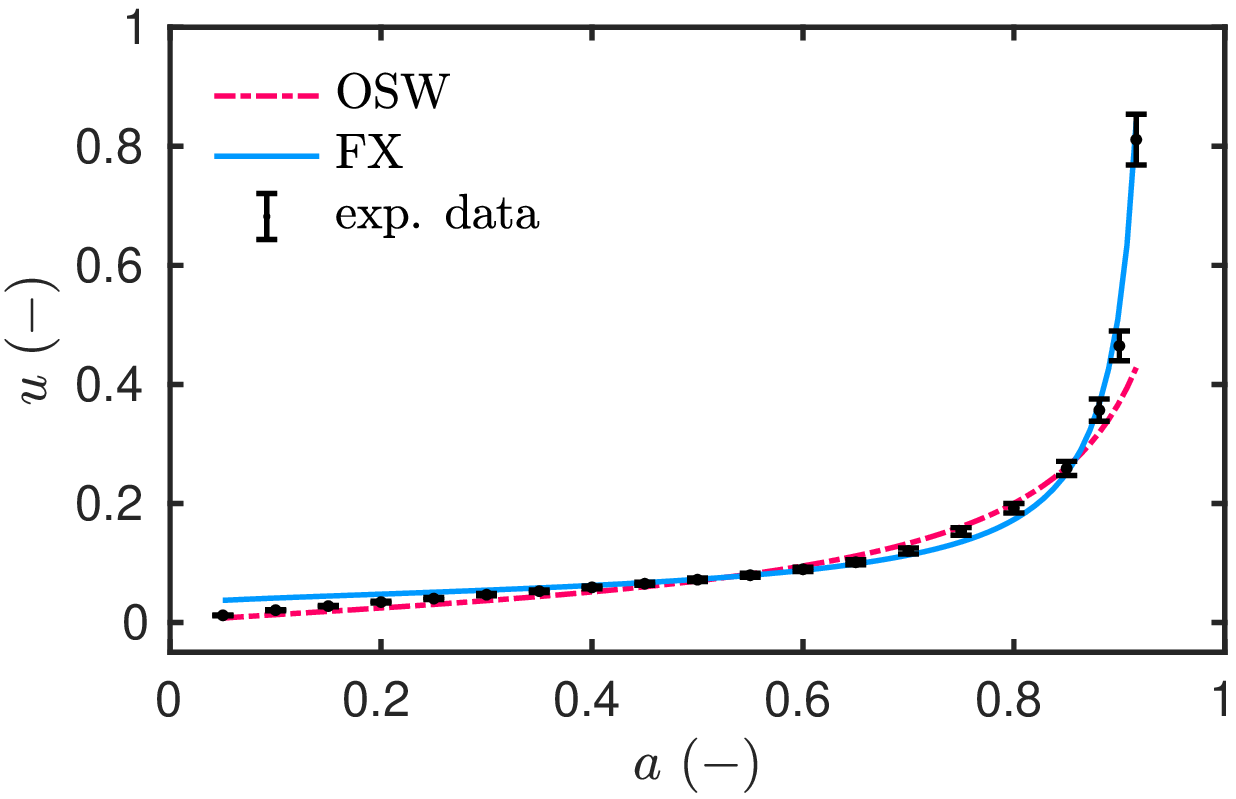}} \hspace{0.2cm}
\subfigure[\label{fig:res_fa_OSW_FX}]{\includegraphics[width=.45\textwidth]{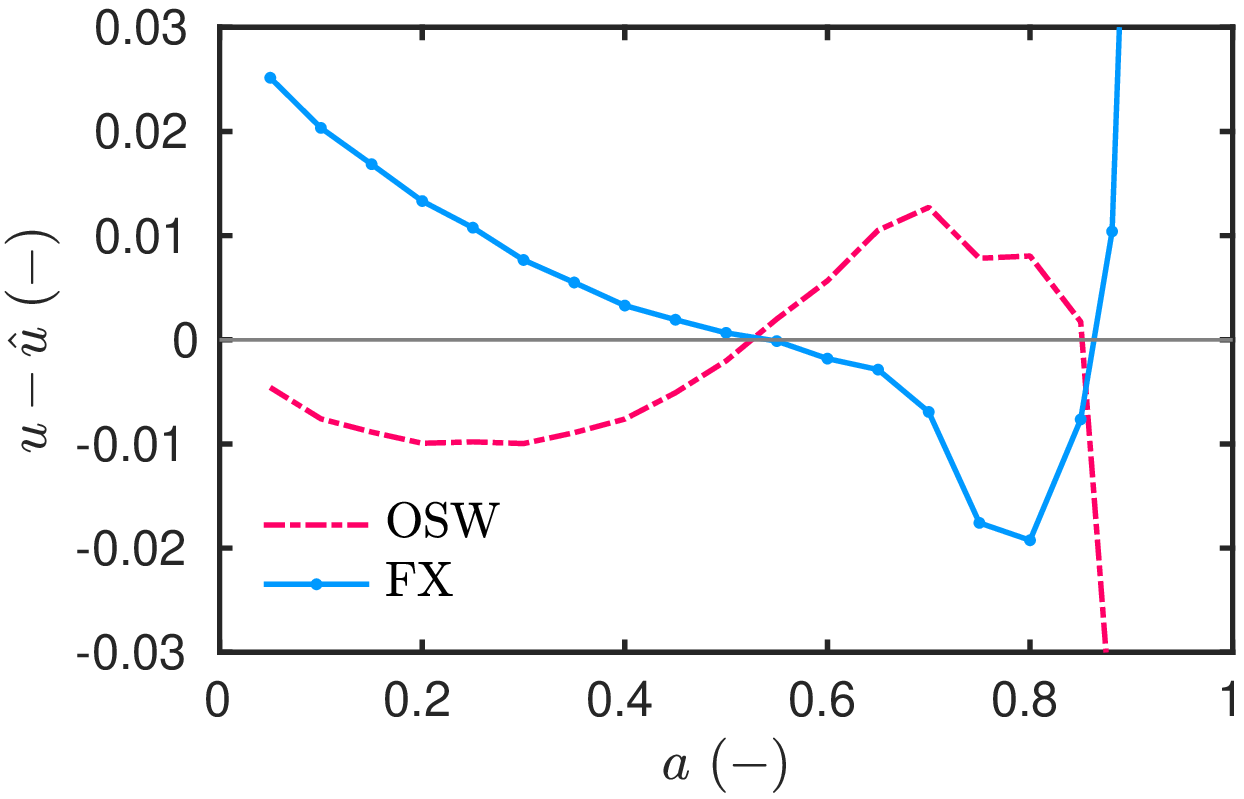}} \\
\subfigure[\label{fig:S_fa_BET_VG}]{\includegraphics[width=.45\textwidth]{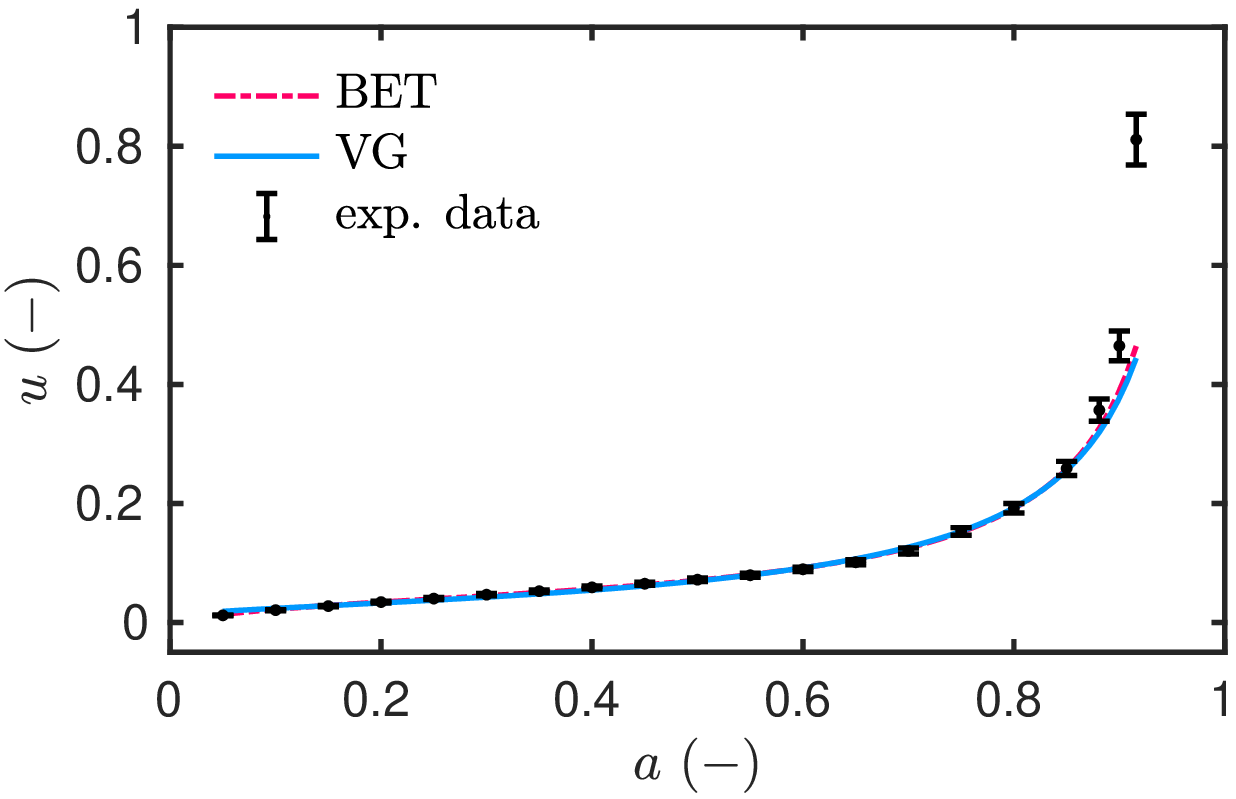}} \hspace{0.2cm}
\subfigure[\label{fig:res_fa_BET_VG}]{\includegraphics[width=.45\textwidth]{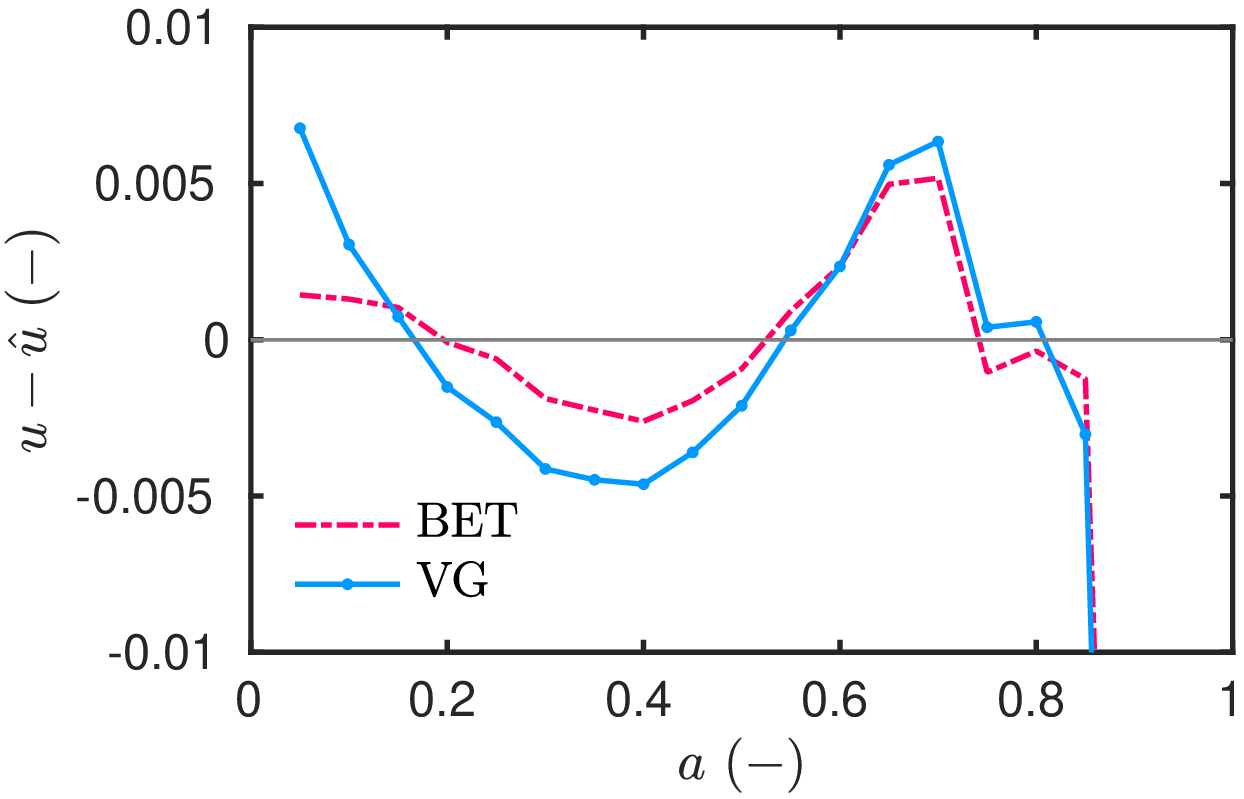}} \\
\subfigure[\label{fig:S_fa_SM_MADS}]{\includegraphics[width=.45\textwidth]{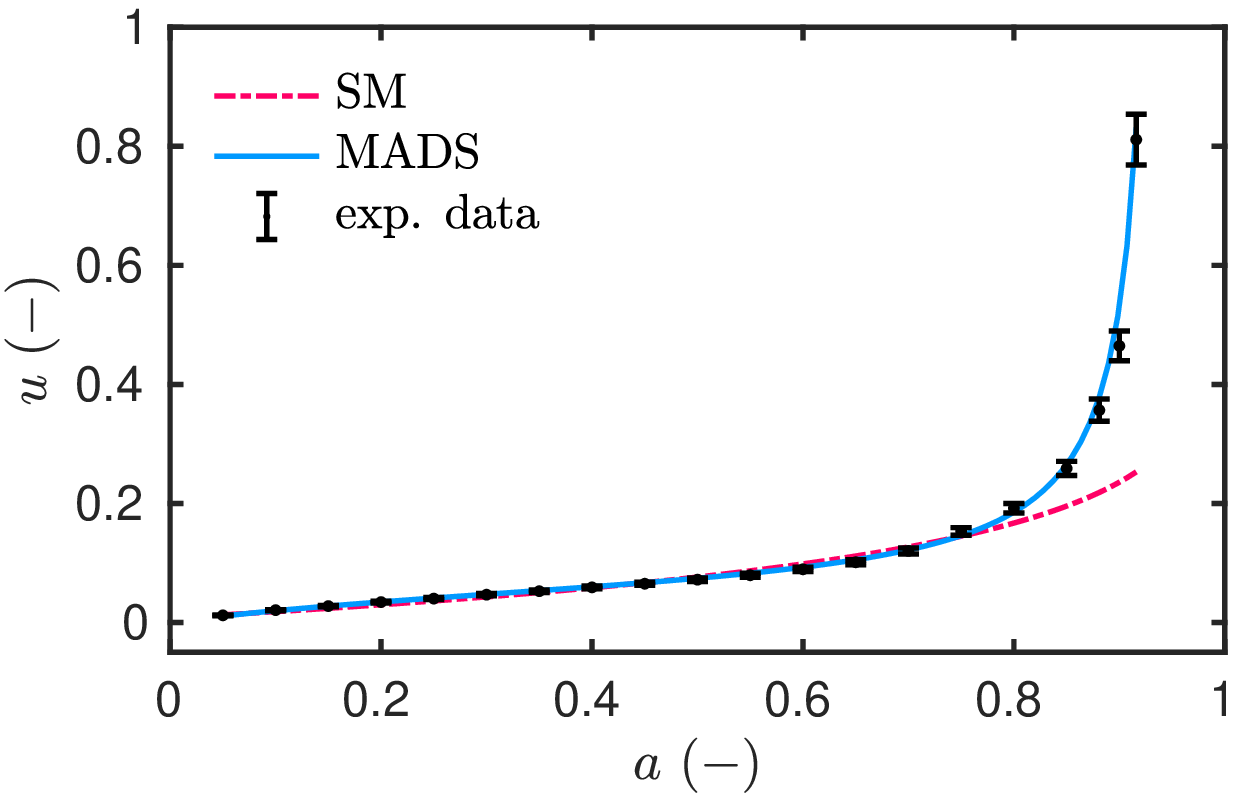}} \hspace{0.2cm}
\subfigure[\label{fig:res_fa_SM_MADS}]{\includegraphics[width=.45\textwidth]{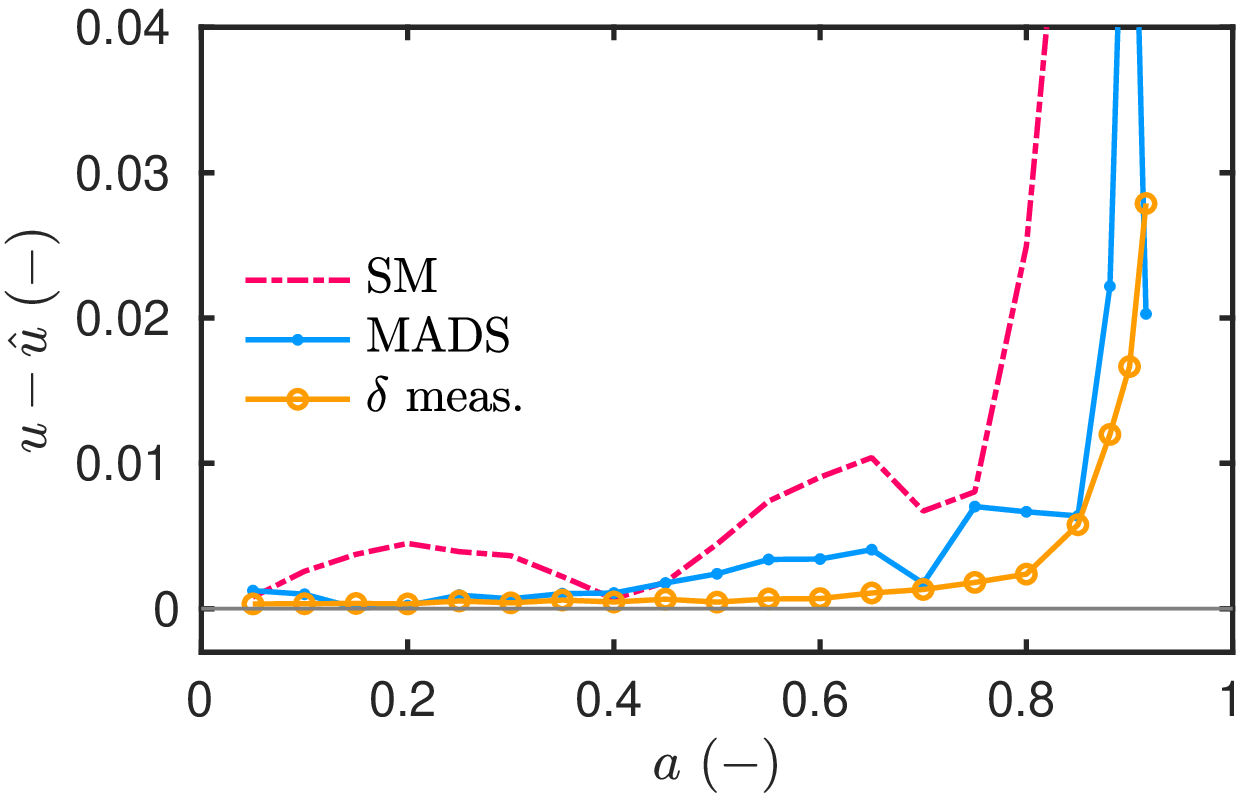}}
\caption{Comparison between the model predictions computed with the estimated parameters and the experimental observations \emph{(a,c,e,g)} and residual between both \emph{(b,d,f,h)}.}
\end{figure}

First, \textsc{Gau}\ss ~algorithm is employed in the least squares sense to solve the parameter estimation problem. The estimated parameters and their estimated uncertainties are reported in Table~\ref{tab:pep_results}. Figures~\ref{fig:S_fa_GAB_BET}, \ref{fig:S_fa_OSW_FX}, \ref{fig:S_fa_BET_VG} and \ref{fig:S_fa_SM_MADS} compare the prediction of the models computed with the estimated parameters and the experimental observations. Figures~\ref{fig:res_fa_GAB_BET}, \ref{fig:res_fa_OSW_FX}, \ref{fig:res_fa_BET_VG} and \ref{fig:res_fa_SM_MADS} present the residuals between computations and observations. Globally, the discrepancies between the model and measurements are relatively low for all models in the so-called hygroscopic state $a \, \leqslant \, 0.80\,$. At high water activity $a \, \geqslant \, 0.80\,$, the discrepancies increase for almost all models, except the MADS one. This can be clearly noticed from the analysis of the residuals. The SM and MADS model residuals have a particular pattern which does not vary around zero. As indicated in Figure~\ref{fig:res_fa_SM_MADS} (and \ref{fig:Sigma_fa}), the pattern is similar to the standard deviation $\delta$ of the measurements. Looking at the distance presented in Table~\ref{tab:pep_results}, the models closest to the experimental observations are the MADS and FX ones. However, for the latter, as shown in Figure~\ref{fig:res_fa_OSW_FX}, the discrepancy is relatively high for $a \, \leqslant \, 0.2\,$. In addition, the model has four parameters to be estimated, which increases the complexity of the estimation problem.

Thus, in general, we note that the parameters could be estimated and the measurements could be accurately predicted with all models.  Nevertheless, the reliability of the models may be discussed looking at the relative estimator error $\eta$ in Table~\ref{tab:pep_results}. Indeed, some parameters are estimated with a very high error estimator. Namely, the parameters $p_{\,3}$ for the GAB model, $p_{\,1}$ for the TRM model, $\bigl(\,p_{\,1}\,,\,p_{\,2}\,,\,p_{\,3}\,,\,p_{\,4}\,\bigr)$ for the FX model, $p_{\,2}$ for the BET model, $\bigl(\,p_{\,1}\,,\,p_{\,2}\,\bigr)$ for the VG model and $p_{\,1}$ for the SM model are estimated with a very low accuracy. For the models FX and VG, the error estimator is higher or equal to half of the standard deviation of the \emph{a priori} uniform distribution. These results are consistent with the ones of the primary identifiability. In other words, parameters with a very low influence on the model prediction correspond to the ones with a high error estimator. For instance, for the TRM model, a bad primary identifiability is observed for parameter $p_{\,1}\,$. Indeed, the estimation of parameter $p_{\,1}$ is very inaccurate with almost $70\%$ of relative error. On the contrary, parameters $p_{\,2}$ and $p_{\,3}$ have good primary identifiability. Consequently, their error estimator is better, around $10\%\,$. 


\begin{table}
\centering
\caption{Results of the parameter estimation problem for each model.}
\label{tab:pep_results}
\setlength{\extrarowheight}{.5em}
\begin{tabular}[l]{@{} c cccc c cccc cc}
\hline
\hline
\multirow{2}{*}{\textit{Models}}
& \multicolumn{4}{c}{\textit{Estimated parameter} } 
&
& \multicolumn{4}{c}{\textit{Relative error estimator} $\eta$} 
&
& \multirow{2}{*}{\textit{Distance} $d_{\,m}\,\bigl(\,\p \,,\,  \hat{u} \,\bigr)$} \\
& $p_{\,1}^{\,\circ}$
& $p_{\,2}^{\,\circ}$
& $p_{\,3}^{\,\circ}$
& $p_{\,4}^{\,\circ}$ 
&
& $p_{\,1}^{\,\circ}$
& $p_{\,2}^{\,\circ}$
& $p_{\,3}^{\,\circ}$
& $p_{\,4}^{\,\circ}$ 
&
&  \\
\hline 
GAB
& $0.035$
& $1.027$
& $15.33$
& -
&
& $0.12$
& $0.04$
& $0.64$
& -
&
& $0.05$ \\
TRM
& $1.156$ 
& $1.383$ 
& $2.165$ 
& -
&
& $0.67$
& $0.15$
& $0.10$
& -
&
& $0.11$ \\
OSW
& $0.069$
& $0.75$
& -
& -
&
& $0.05$
& $0.09$
& -
& -
&
& $0.15$ \\
FX
& $7.98$ 
& $18.39$ 
& $9.586$ 
& $1.471$ 
&
& $>\,1$
& $>\,1$
& $>\,1$
& $0.48$
&
& $0.007$ \\
BET
& $0.04$
& $9.18$ 
& -
& -
&
& $0.07$
& $0.44$
& -
& -
&
& $0.12$ \\
VG
& $51.33$
& $2299$ 
& $1.895$
& -
&
& $>\,1$
& $>\,1$
& $0.12$
& -
&
& $0.14$ \\
SM
& $0.0078$
& $0.099$ 
& -
& -
&
& $0.53$
& $0.09$
& -
& -
&
& $0.38$ \\
MADS
& $-0.76$
& $2.47$
& -
& -
&
& $0.04$
& $0.04$
& -
& -
&
& $0.005$ \\
\hline
\hline
\end{tabular}
\end{table}

\subsubsection{Model selection}

\begin{figure}
\centering
\subfigure[\label{fig:Tol_fpop}]{\includegraphics[width=.45\textwidth]{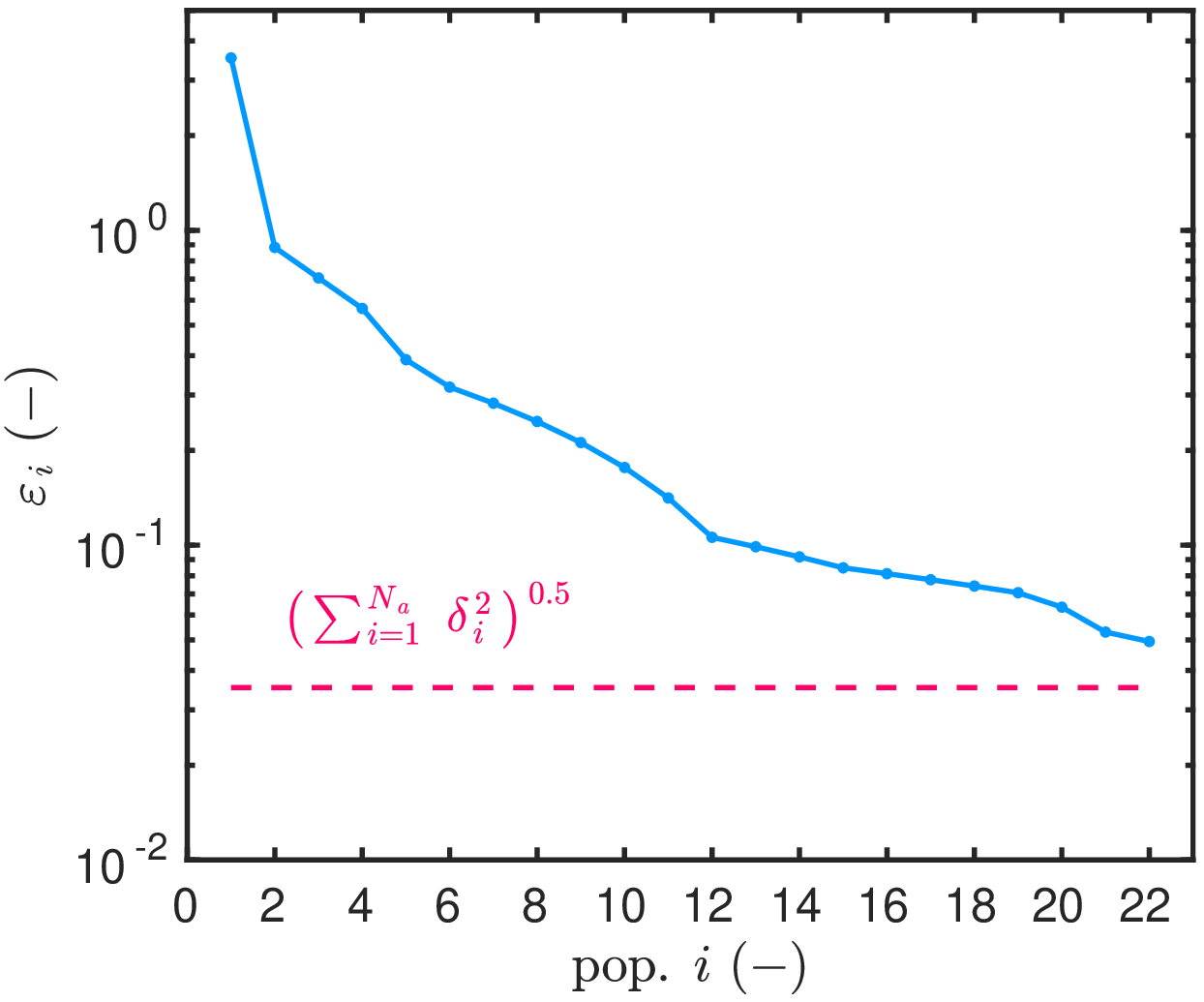}} \hspace{0.2cm}
\subfigure[\label{fig:Taxa_fpop}]{\includegraphics[width=.45\textwidth]{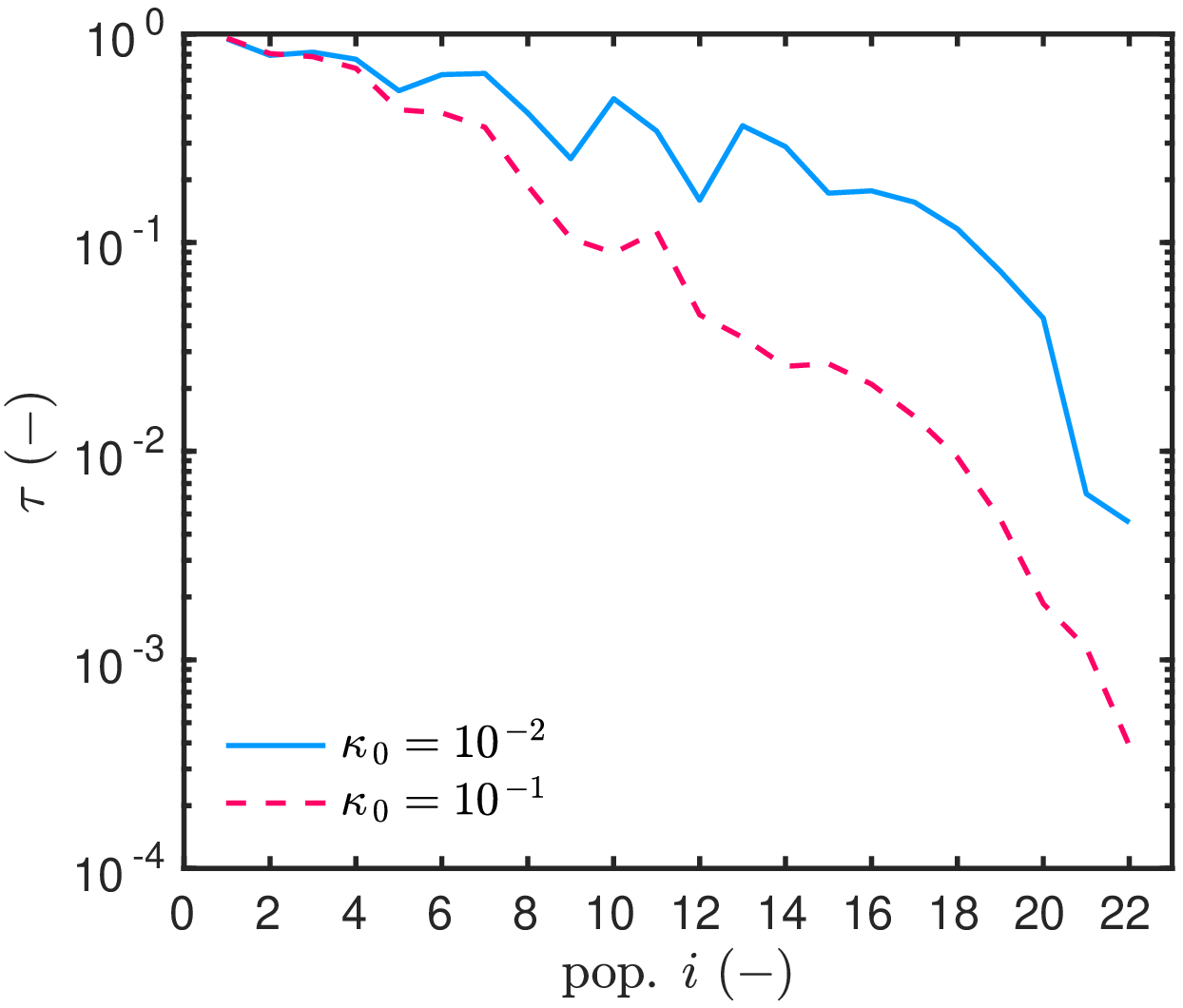}} 
\caption{Variation of the tolerance \emph{(a)} and the acceptance rate \emph{(b)} according to the population.}
\end{figure}

Now, the ABC algorithm is used for model selection and model calibration (estimation of the model parameters), among the eight competing ones described above, for a kernel $\kappa_{\,0} \egal 0.01\,$. A number of $22$ populations are chosen as illustrated in Figure~\ref{fig:Tol_fpop}. The tolerance is decreasing with the number of population respecting the \textsc{Morozov}'s discrepancy principle at the final population. The last tolerance scales with the square roots of the sum of the uncertainties for each measurement $ \displaystyle \varepsilon_{\,22} \egal 1.4 \cdot \sqrt{ \sum_{i =1}^{N_{\,a}} \ \delta_{\,i}^{\,2} } \,$, $N_{\,a}$ being the total number of measurements. A total number $N_{\,\nu} \egal 4000$ particles is chosen. Figure~\ref{fig:Taxa_fpop} shows the variation of the acceptance rate according to each population. It decreases with the population. For the last population, the acceptance rate is $0.4 \%\,$. 

\begin{figure}
\centering
\subfigure[\label{fig:Np_fpop_k1} $\kappa_{\,0} \egal 0.01$]{\includegraphics[width=.9\textwidth]{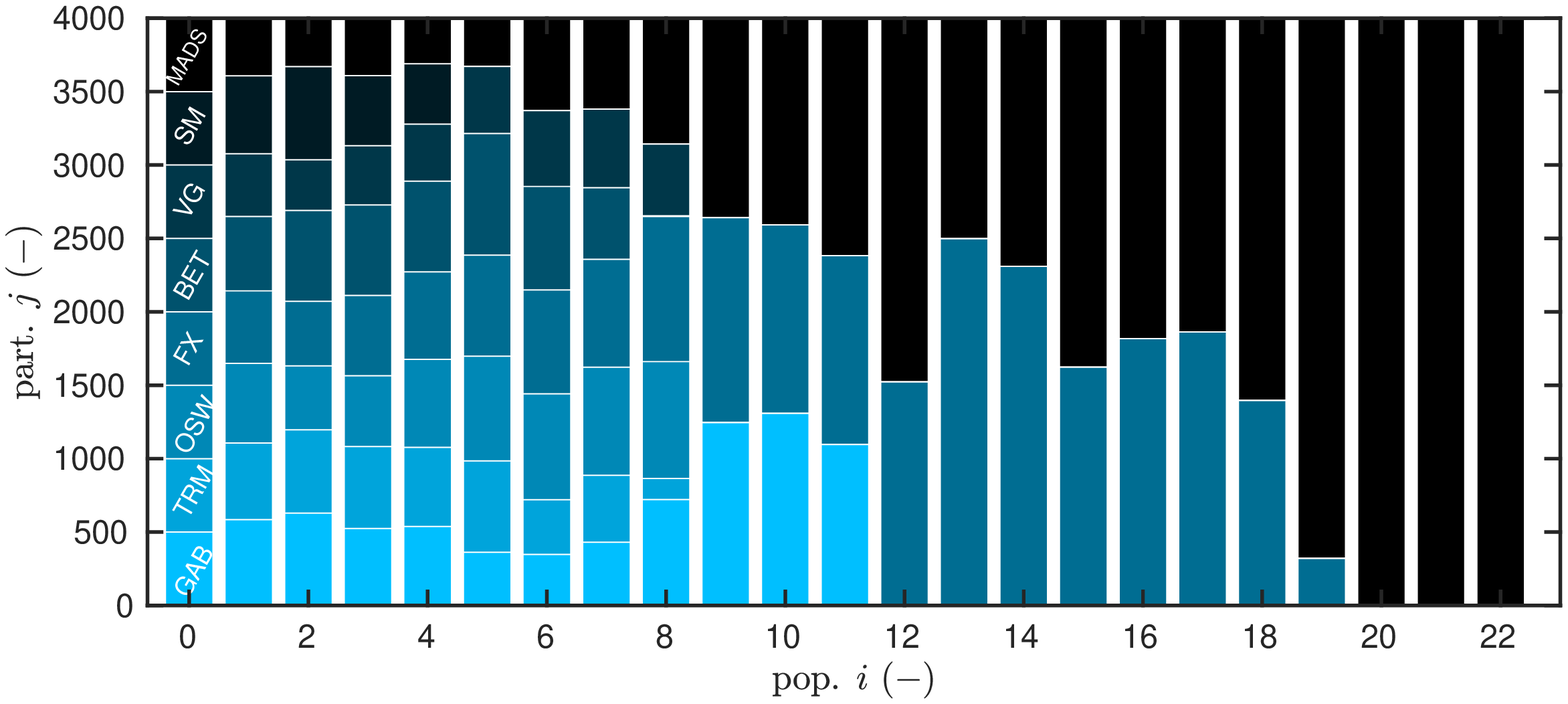}} \\
\subfigure[\label{fig:Np_fpop_k2} $\kappa_{\,0} \egal 0.1$]{\includegraphics[width=.9\textwidth]{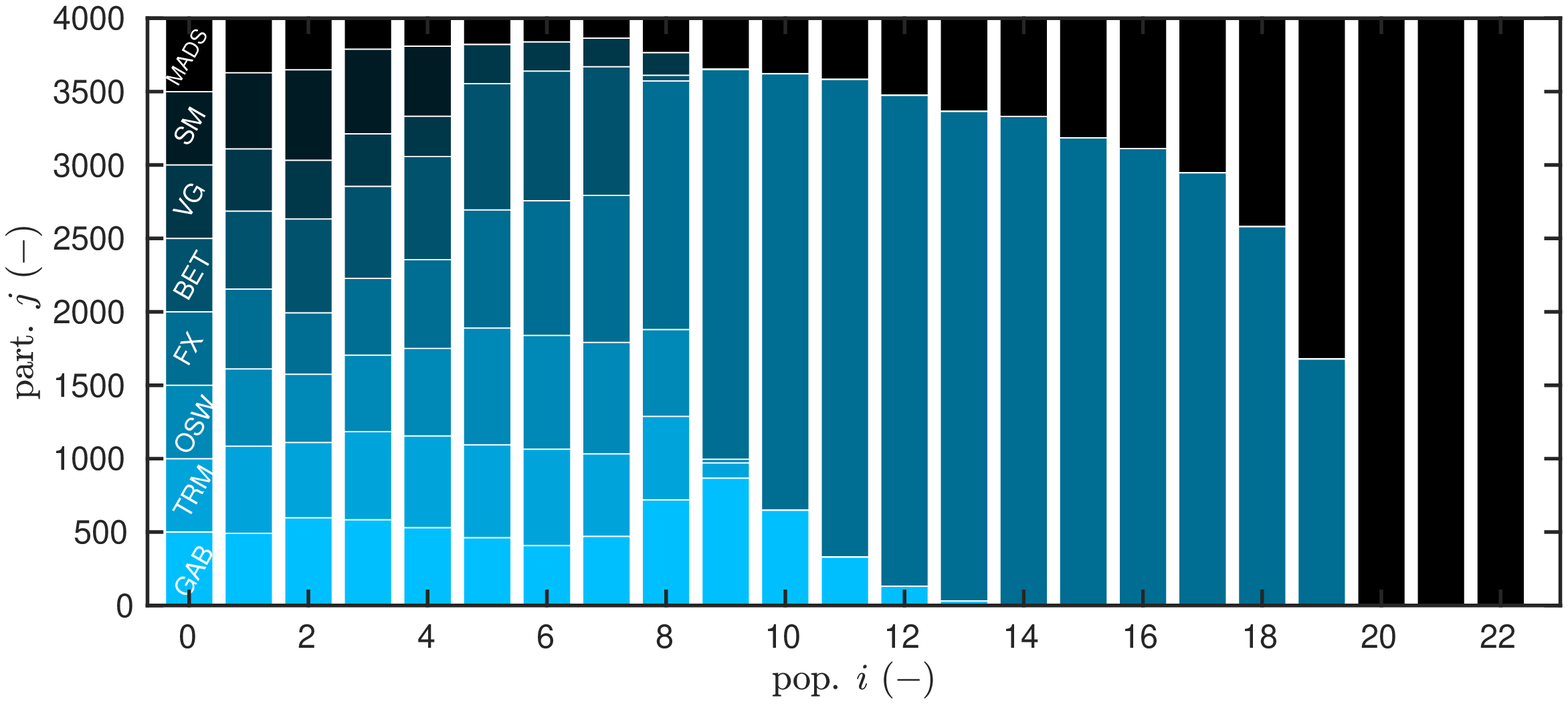}} 
\caption{Variation of the model selection according to the population for different kernel.}
\end{figure}

Figure~\ref{fig:Np_fpop_k1} shows the model selection through the $22$ populations. Initially, the eight models have an equal number of particles. Until population $4\,$, there is no strong selection among the models. The SM model becomes less selected since it does not succeed in representing the phenomena for high water activity, as reported in Figure~\ref{fig:S_fa_SM_MADS}. At population $5\,$, no particles validate the distance test for this model with the candidate parameters. Thus, at population $5\,$, the model SM is no longer selected. At population $9\,$, it can be noticed that only $3$ models are still competing, namely the MADS, the FX and the GAB. At the final population, the MADS model is the only one satisfying the distance test for the lowest tolerance. 

\begin{figure}
\centering
\subfigure[\label{fig:p1_fp2_OSW_MADS}]{\includegraphics[width=.90\textwidth]{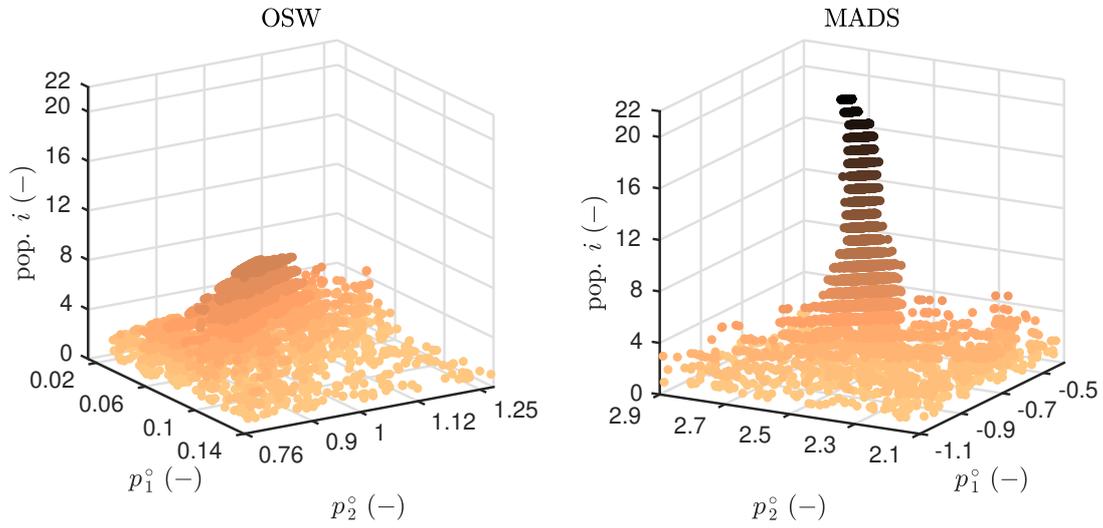}} \\
\subfigure[\label{fig:pcirc_GAB_MADS_fpop}]{\includegraphics[width=.90\textwidth]{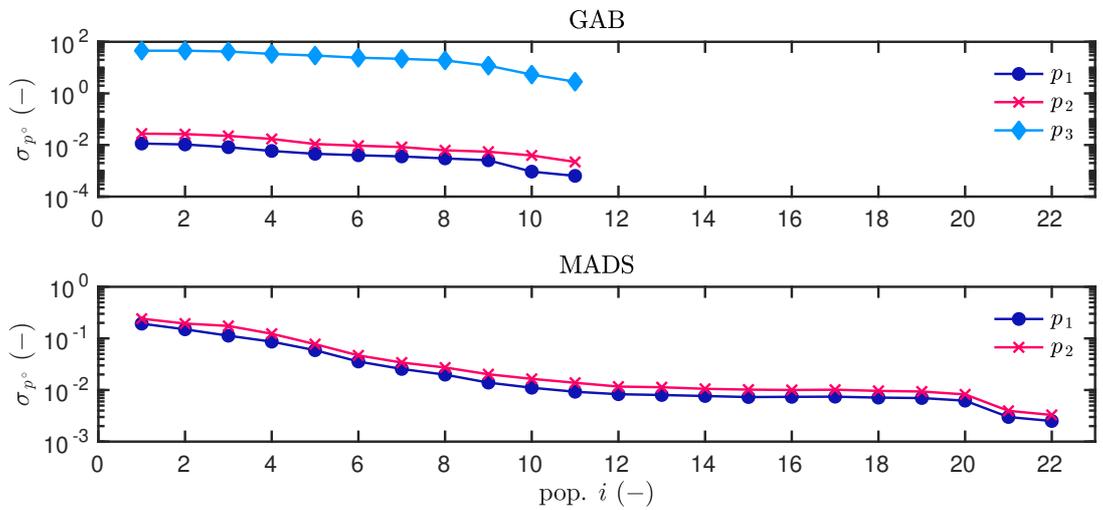}}
\caption{Variation of the estimated parameters \emph{(a)} and their standard deviation \emph{(b)} according to the population.}
\end{figure}

The model selection can also be discussed by analysing the evolution of estimated parameters according to the population presented in Figures~\ref{fig:p1_fp2_OSW_MADS} and \ref{fig:pcirc_GAB_MADS_fpop}. Figure~\ref{fig:p1_fp2_OSW_MADS} shows the dissemination of the estimated parameters in the plan $\Omega_{\,p_{\,1}} \, \times \, \Omega_{\,p_{\,2}} $ according to the population, for both models OSW and MADS. For the first population, the estimated parameters are scattered. At population $8\,$, for the OSW model, the estimated parameters still exhibit a large variability. For the MADS model, the parameters are already concentrated in a narrow region, corresponding to the final estimated parameters of the model. This analysis is consistent with the evolution of the standard deviation of the estimated parameters according to the population. Figure~\ref{fig:pcirc_GAB_MADS_fpop} shows that the GAB model is not selected anymore after population $11\,$. Indeed, the standard deviation of the parameters estimated for this model is large. The algorithm did not find new candidate parameters that validate the distance test of the GAB model. 

\begin{figure}
\centering
\subfigure[\label{fig:hist_P1}]{\includegraphics[width=.45\textwidth]{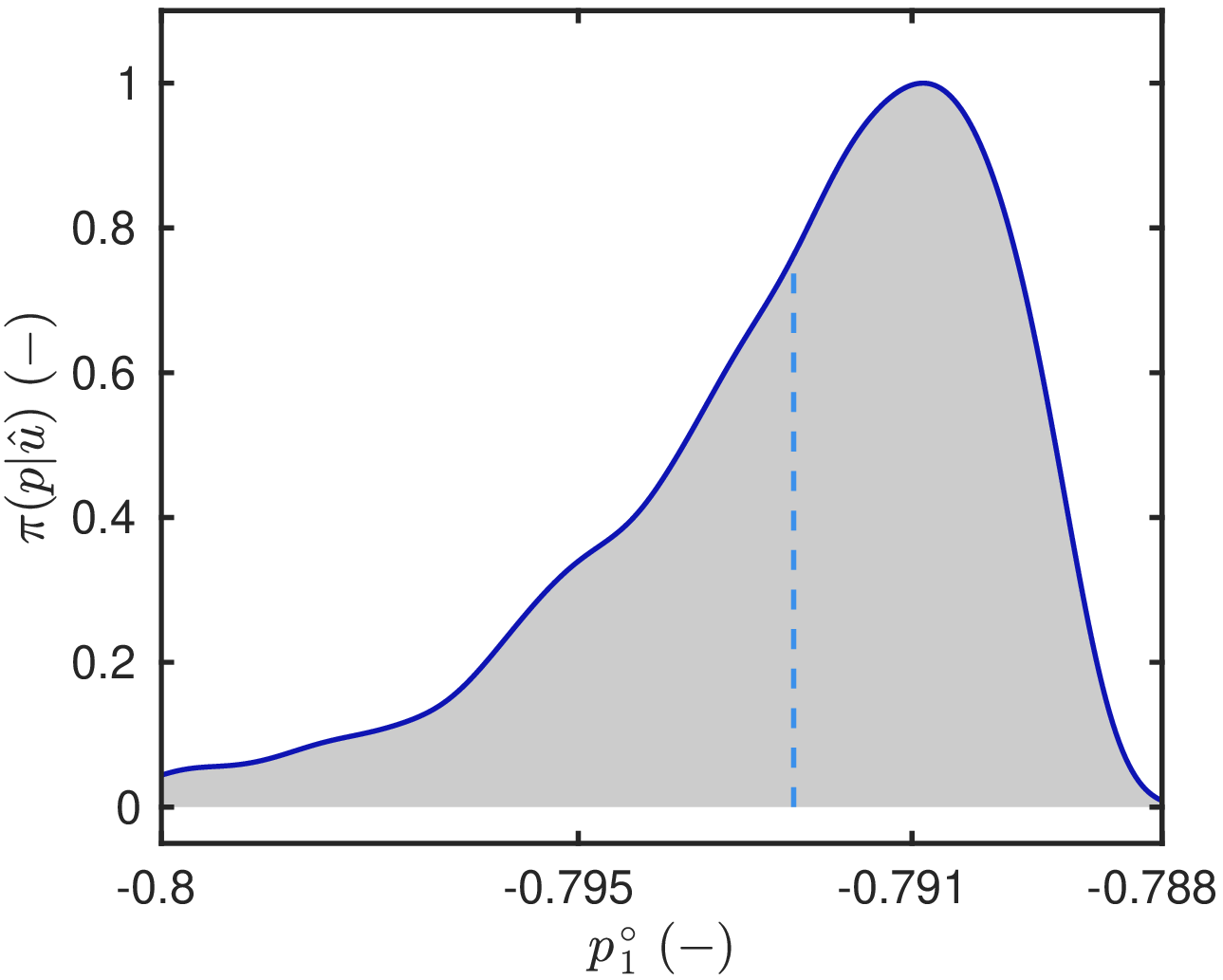}} \hspace{0.2cm}
\subfigure[\label{fig:hist_P2}]{\includegraphics[width=.45\textwidth]{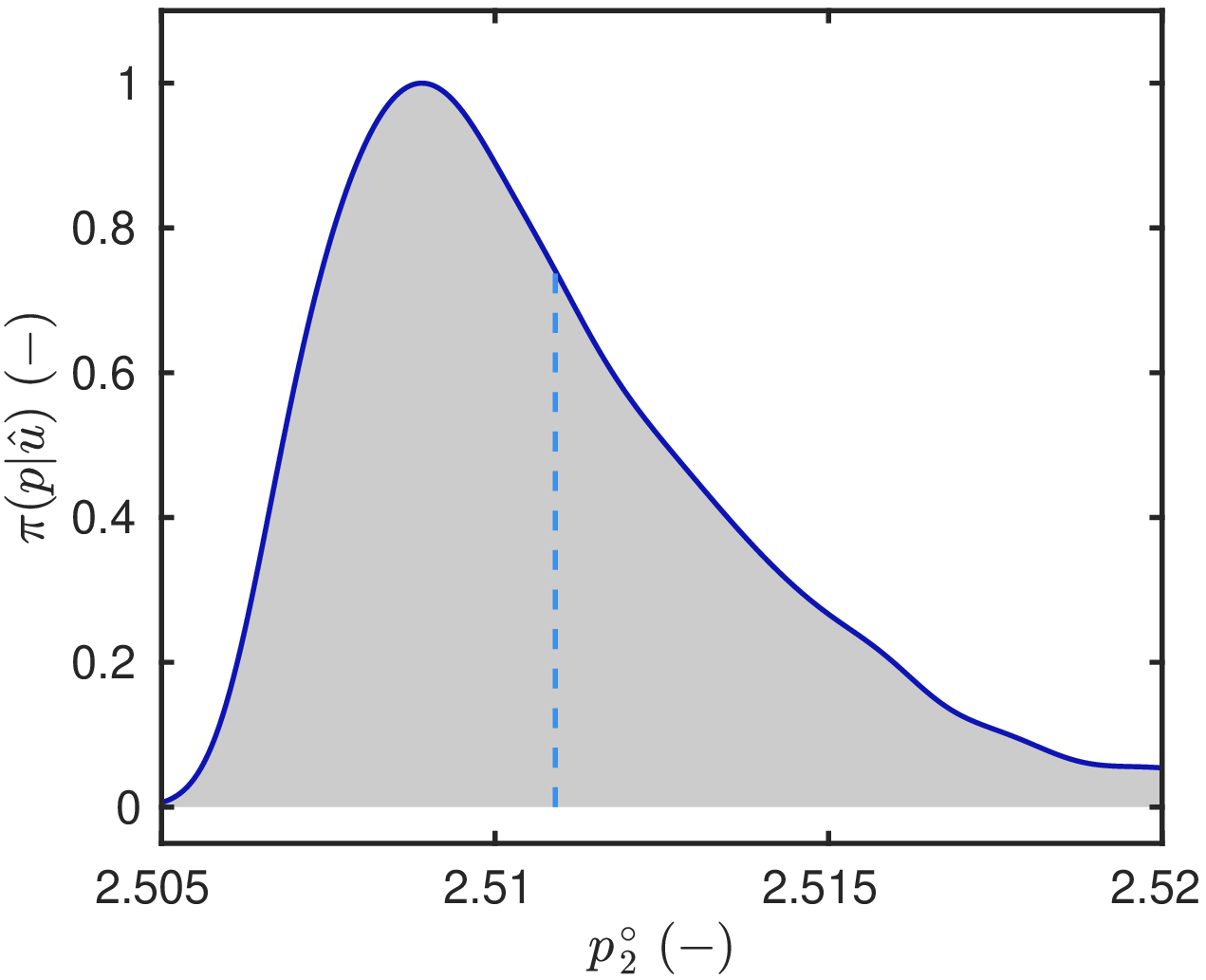}} 
\caption{Normalized posterior distribution of the estimated parameters $p_{\,1}^{\,\circ}$ \emph{(a)} and $p_{\,2}^{\,\circ}$ \emph{(b)} for the MADS.}
\label{fig:hist_Pest}
\end{figure}

At the final population, the parameters are estimated for the selected model. Figure~\ref{fig:hist_Pest} gives the posterior distribution of the parameters $p_{\,1}$ and $p_{\,2}$ of the MADS model. It can be noticed that it corresponds to the one estimated in previous section, using the least estimator algorithm, in Table~\ref{tab:pep_results}. In addition, the standard deviation of posterior distributions is very low indicating an accurate estimation.

To confirm the results, the model selection is performed for another kernel $\kappa_{\,0} \egal 0.1\,$, while keeping $22$ populations. As noted in Figure~\ref{fig:Taxa_fpop}, the acceptance rate $\tau$ decreases when $\kappa_{\,0}$ increases. Thus, the computational time of the algorithm rises significantly (from $15 \ \mathsf{min}$ for $\kappa_{\,0} \egal 0.01$ to $30 \ \mathsf{min}$ for $\kappa_{\,0} \egal 0.1$ in the \texttt{Matlab\texttrademark} environment with a computer equipped with \texttt{Intel} i$7$ CPU and $32$ GB of RAM). The model selection among the $22$ populations is shown in Figure~\ref{fig:Np_fpop_k2}. The results are unchanged, while the MADS model being selected. It is noted that additional simulations have been carried for higher kernel $\kappa_{\,0} \egal 1\,$ and $400$ particles, also resulting in the selection of the MADS model. However, the acceptance rate was too low indicating a bad choice for the kernel parameter.

\section{Conclusion}
\label{sec:conclusion}

The sorption curve is an essential property for wood materials with the objective of modelling the interaction of wood with ambient moist air. This paper proposes to investigate the reliability of eight models through their robustness to the identified parameters. Seven have been proposed in the literature with various examples of applications. The last one is proposed based on the general shape of moisture sorption curve for porous building material. Experimental measurements are taken for a wood fibre material. Using a DVS equipment, the moisture content is obtained according to several levels of water activity.

Using the experimental observations, the reliability of the models is discussed in Section~\ref{sec:reliability_models} through the accuracy of the parameter estimation. The so-called primary identifiability of each parameter is discussed. It investigates if the parameters sufficiently influence the output of the model to be estimated with accuracy. For this, a continuous derivative-based approach is adopted using the sensitivity function of the model. A global sensitivity metric is computed for each parameter. Seven models have a low primary identifiability for at least one parameter. In other words, one parameter is not influencing the model sufficiently to be estimated with accuracy. These results are confirmed when solving the parameter estimation problem in Section~\ref{sec:parameter_estimation_problem}. For all models, a set of parameters can be identified to represent accurately the moisture sorption curve. However, the parameter with low primary identifiability are retrieved with a large error estimator. Last, an ABC algorithm is used for simultaneously model selection and model calibration, among eight competing models. The proposed model appears to have the best reliability based on the distance between measurements and estimation. The so-called thermodynamic \citep{Merakeb_2009} and \textsc{Feng}--\textsc{Xing} \citep{Fredlund_1994} models are also reliable candidates. The first has a high parametric complexity since it is composed of four parameters.

Future works should focus on the hysteresis effects. It should be taken into account since it has a significant influence on the precision of the numerical predictions of heat and mass transfer \citep{Zhang_2016a}.

\section*{Nomenclature and symbols}

\begin{tabular}{|cll|}
\hline
\multicolumn{3}{|c|}{\emph{Physical parameters}} \\ \hline
\multicolumn{3}{|c|}{Latin letters} \\ 
$a\,,\,\hat{a}$ & water activity & $\unit{-}$ \\
$K$ & slope of the sorption model & $\unit{-}$ \\
$m\,,m_{\,0}$ & mass & $\unit{kg}$ \\
$p$ & parameter of a model & $\unit{-}$ \\
$R_{\,1}$ & water vapor gas constant & $\unit{J\,.\,kg^{\,-1}\,.\,K^{\,-1}}$ \\
$T$ & temperature & $\unit{K}$ \\
$u\,,u_{\,0}\,,\,\hat{u}\,,u_{\,m}$ & moisture content dry basis & $\unit{-}$ \\\hline
\multicolumn{3}{|c|}{Greek letters} \\
$\delta$ & measurement uncertainty  & $\unit{-}$ \\
$\delta_{\,\sim}$ & random component of measurement uncertainty  & $\unit{-}$ \\
$\delta_{\,\sigma}$ & systematic component of measurement uncertainty   & $\unit{-}$ \\
$\Psi$ & capillary pressure & $\unit{Pa}$ \\
$\rho$ & material dry density & $\unit{kg\,.\,m^{\,-3}}$ \\
$\rho_{\,2}$ & liquid water specific mass & $\unit{kg\,.\,m^{\,-3}}$ \\
\hline
\end{tabular}

\hspace{2cm}
\bigskip

\begin{tabular}{|cll|}
\hline
\multicolumn{3}{|c|}{\emph{Mathematical notations}} \\
\hline
\multicolumn{3}{|c|}{Latin letters} \\
$d$ & distance function $i$ & \\
$\mathfrak{D}_{\,i}$ & polynomial of order $i$ & \\
$f$ & sorption model function & \\
$F$ & \textsc{Fisher} matrix & \\
$\mathcal{K}$ & kernel perturbation & \\
$N$ & number of parameters of a model & \\
$N_{\,a}$ & number of measurement point for a population & \\
$N_{\,e}$ & number of carried measurement & \\
$N_{\,\varepsilon}$ & number of populations & \\
$N_{\,\nu}$ & number of particles in each population & \\
$\mathfrak{N}_{\,i}$ & polynomial of order $i$ & \\
$\mathfrak{P}_{\,N}$ & polynomial of order $N$ & \\
$\mathcal{U}$ & uniform distribution & \\
$w$ & weight & \\
\hline
\multicolumn{3}{|c|}{Greek letters} \\
$\alpha_{\,i}$ & polynomial coefficients  & \\
$\beta_{\,i}$ & polynomial coefficients  & \\
$\varepsilon \,,\, \varepsilon_{\,i}$ & tolerance & \\
$\eta_{\,n}$ & relative error estimator  & \\
$\gamma_{\,n}\,,\,\gamma^{\,\intercal}$ & derivative-based sensitivity metric related to parameter $p_{\,n}$ & \\
$\Omega$ & set of elements & \\
$\pi$ & probability distribution & \\
$\nu_{\,n}\,,\,\nu^{\,\intercal}$ & local derivative-based sensitivity metric related to parameter $p_{\,n}$ & \\
$\kappa\,,\, \kappa_{\,0}$ & kernel parameter & \\
$\tau$ & acceptation rate & \\
$\theta_{\,n}$ & sensitivity function related to parameter $p_{\,n}$ & \\
\hline
\multicolumn{3}{|c|}{Subscripts and superscripts} \\
$a$ & related to water activity &  \\
$\apr$ & \emph{a priori} parameter &  \\
$i$ & population indicator &  \\
$j$ & particle indicator &  \\
$m$ & model indicator &  \\
$n$ & parameter indicator &  \\
$p$ & related to a parameter &  \\
$0$ & dry basis property &  \\
$\circ$ & estimated parameter &  \\
$\star \,,\, \star \star$ & sampled parameter in ABC algorithm &  \\
$\wedge$ & measured quantity obtained from experiments &  \\
$\sim $ & different member from the same set (SGI property) &  \\
$\intercal$ & dimensionless metric &  \\
\hline
\end{tabular}

%
%

\section*{Acknowledgments}

The authors acknowledge the Junior Chair Research program ``Building performance assessment, evaluation and enhancement'' from the University of Savoie Mont Blanc in collaboration with The French Atomic and Alternative Energy Center (CEA) and Scientific and Technical Center for Buildings (CSTB). The authors acknowledge the French and Brazilian agencies for their financial support through the project CAPES--COFECUB, as well as the CNPQ of the Brazilian  Ministry of Education and of the Ministry of Science, Technology and Innovation, respectively, for co-funding. The support provided by FAPERJ, agency of the Rio de Janeiro state government is gratefully appreciated. The authors also thank A. Moreau and A. Jumabekova for the productive discussions on the MADS and GAB models, respectively.

\section*{Conflict of interests}

Authors have no conflict of interest relevant to this article.

\bibliographystyle{apalike}
\bibliography{references}

\newpage

\appendix

\section{Structural identifiability of the sorption models}
\label{app:structural_identifiability}

The purpose is to demonstrate that the unknown parameters of the sorption models are identifiable from a theoretical point of view. For this, the SGI property is evaluated for each model. 

\subsection{Property of structural identifiability}

First, the structural identifiability of the unknown parameters should be demonstrated. It aims at stating if the parameters, according to the sorption model, are theoretically identifiable. Using the definition provided in \citep{Walter_1982,Walter_1990}, a parameter $\p \, \in \, \Omega_{\,p\,,\,m}$ is Structurally Globally Identifiable (SGI) in the model $f$ if the following condition is verified:
\begin{align*}
\forall \, a \, \in \, \Omega_{\,a} \,, \qquad f_{\,m}\,\bigl(\,\p\,,\,a \,\bigr) \egal f_{\,m}\, \bigl(\,\tilde{\p}\,,\,a\,\bigr) \, \Longrightarrow \, \p \egal \tilde{\p} \,.
\end{align*}
This property should be evaluated for each unknown parameter of the eight sorption models. 

\subsection{GAB model}

A first set of observable $u$ is obtained for the GAB model with parameters $p \egal \bigl(\,p_{\,1}\,,\,p_{\,2}\,,\,p_{\,3}\,)$. A second one is also hold, denoted as $\tilde{u}\,$, for the parameters $\tilde{p} \egal \bigl(\,\tilde{p}_{\,1}\,,\,\tilde{p}_{\,2}\,,\,\tilde{p}_{\,3}\,)\,$. Using the model definition Eq.~\eqref{eq:GAB_model} and a symbolic computing environment, the operation $u \moins \tilde{u}$ is performed to obtain an algebraic fraction of the form:
\begin{align*}
F\,(\,a\,) \egal \frac{\mathfrak{N}_{\,3}(\,a\,)}{\mathfrak{D}_{\,4}(\,a\,)}\,, 
\end{align*}
where $\mathfrak{N}_{\,3}$ and $\mathfrak{D}_{\,4}$ are third and fourth order polynomials of $a\,$, respectively:
\begin{align*}
\mathfrak{N}_{\,3} \egal \sum_{k\,=\,0}^3 \, \alpha_{\,i} \ a^{\,i} \,, \qquad \mathfrak{D}_{\,4} \egal \sum_{k\,=\,0}^4 \, \beta_{\,i} \ a^{\,i} \,.
\end{align*}
Thus, we have $u \moins \tilde{u} \, \equiv \, \bold{0}\,$, $ \forall a \, \in \, \Omega_{\,a}\,$, if and only if the coefficients of the polynomial $N_{\,3}$ are equal to zero:
\begin{align}
\label{eq:cond_GAB}
\alpha_{\,k} \egal 0 \,, \qquad \forall \, k \, \in \, \bigl\{\,0\,,\,1\,,\,2\,,\,3 \,\bigr\}\,. 
\end{align}
It can be demonstrated that Eq.~\eqref{eq:cond_GAB} leads to $p_{\,n} \, \equiv \, \tilde{p}_{\,n} \,, \forall \, n \, \in \,  \bigl\{\,1\,,\,2\,,\,3 \,\bigr\} \,$. Therefore, the parameters $p \egal \bigl(\,p_{\,1}\,,\,p_{\,2}\,,\,p_{\,3}\,)$ are SGI. 


\subsection{TRM model}

Two sets of observables $u$ and $\tilde{u}$ are admitted for the TRM model with parameters $p$ and $\tilde{p}\,$, respectively. Using the model definition~\eqref{eq:TRM_model}, the equality $u \, \equiv \, \tilde{u}$ leads to:
\begin{align}
\label{eq:cond_TRM}
\frac{p_{\,1}}{\tilde{p}_{\,1}}
 \ \exp \Bigl(\, p_{\,2} \ \ln \bigl(\, a \,\bigr) \cdot \exp \bigl(\, p_{\,3} \ a \,\bigr)
\moins  \tilde{p}_{\,2} \ \ln \bigl(\, a \,\bigr) \ \exp \bigl(\, \tilde{p}_{\,3} \ a \,\bigr)
\,\Bigr)
 \egal \bold{1} \,.
\end{align}
If we assume $p_{\,1} \, \equiv \, \tilde{p}_{\,1}\,$, and thus $p_{\,1}$ being SGI, then from Eq.~\eqref{eq:cond_TRM} we must have:
\begin{align*}
p_{\,2} \ \ln \bigl(\, a \,\bigr) \ \exp \bigl(\, p_{\,3} \ a \,\bigr)
\moins  \tilde{p}_{\,2} \ \ln \bigl(\, a \,\bigr) \ \exp \bigl(\, \tilde{p}_{\,3} \ a \,\bigr) \, \equiv \, \bold{0} \,,
\end{align*}
which can be rewritten as:
\begin{align*}
\frac{p_{\,2}}{\tilde{p}_{\,2}} \ \exp \bigl(\, p_{\,3} \ a \moins \tilde{p}_{\,3} \ a\,\bigr) \, \equiv \, \bold{1} \,.
\end{align*}
Again, if we assume $p_{\,2} \, \equiv \, \tilde{p}_{\,2}\,$, and thus $p_{\,2}$ being SGI, it follows that:
\begin{align*}
p_{\,3} \, \equiv \, \tilde{p}_{\,3} \,,
\end{align*}
and parameter $p_{\,3}$ is SGI. By reciprocity, the parameters $p_{\,1}$ and $p_{\,2}$ are SGI.

\subsection{OSW model}

It is assumed two sets of observable $u$ and $\tilde{u}$ of the OSW model for the same water activity $a$ and with parameters $p$ and $\tilde{p}\,$. The model definition gives for the $u \, \equiv \, \tilde{u}\,$:
\begin{align*}
\frac{p_{\,1}}{\tilde{p}_{\,1}} \ \biggl(\, \frac{a}{1 \moins a} \,\biggr)^{\,p_{\,2} \moins \tilde{p}_{\,2}} \, \equiv \, \bold{1} \,.
\end{align*}
This equality is true if and only $p_{\,1} \, \equiv \, \tilde{p}_{\,1}$ and $p_{\,2} \, \equiv \, \tilde{p}_{\,2}\,$. Thus, parameters $p_{\,1}$ and $p_{\,2}$ are SGI for the OSW model.

\subsection{FX model}

For the FX model, two sets of observable  $u$ and $\tilde{u}$ are obtained for the parameters $p$ and $\tilde{p}\,$, respectively. The equality $u \, \equiv \, \tilde{u}$ gives using the model definition:
\begin{align*}
\frac{p_{\,1}}{\tilde{p}_{\,1}} \ \frac{\Biggl[\, \ln \Biggl(\,\mathrm{e} 
\plus \biggl(\, - \, p_{\,2} \ \ln (\,a\,)\,\biggr)^{\,p_{\,3}} \,\Biggr)
\,\Biggr]^{\,-\,p_{\,4}}}
{\Biggl[\, \ln \Biggl(\,\mathrm{e} 
\plus \biggl(\, - \, \tilde{p}_{\,2} \ \ln (\,a\,)\,\biggr)^{\,\tilde{p}_{\,3}} \,\Biggr)
\,\Biggr]^{\,-\,\tilde{p}_{\,4}}} \, \equiv \, \bold{1} \,.
\end{align*}
If we assumed that both parameter $p_{\,1}$ and $p_{\,4}$ are SGI, and thus $p_{\,1} \, \equiv \, \tilde{p}_{\,1}$ and $p_{\,4} \, \equiv \, \tilde{p}_{\,4}\,$, it follows that:
\begin{align*}
\frac{\ln \Biggl(\,\mathrm{e} 
\plus \biggl(\, - \, p_{\,2} \ \ln (\,a\,)\,\biggr)^{\,p_{\,3}} \,\Biggr)
}
{\ln \Biggl(\,\mathrm{e} 
\plus \biggl(\, - \, \tilde{p}_{\,2} \ \ln (\,a\,)\,\biggr)^{\,\tilde{p}_{\,3}} \,\Biggr)
} \, \equiv \, \bold{1} \,,
\end{align*}
and thus 
\begin{align*}
\mathrm{e} 
\plus \biggl(\, - \, p_{\,2} \ \ln (\,a\,)\,\biggr)^{\,p_{\,3}}
\, \equiv \,
\mathrm{e} 
\plus \biggl(\, - \, \tilde{p}_{\,2} \ \ln (\,a\,)\,\biggr)^{\,\tilde{p}_{\,3}} \,.
\end{align*}
Thus, if parameter $p_{\,3}$ is assumed SGI, we obtain that $p_{\,2} \, \equiv \, \tilde{p}_{\,2}\,$. So parameter $p_{\,2}$ is also SGI. In brief, the equality $u \, \equiv \, \tilde{u}$ admits the solution where parameters $p_{\,1}\,$, $p_{\,2}\,$, $p_{\,3}$ and $p_{\,4}$ are SGI. 

\subsection{BET model}

Two sets of observable $u$ and $\tilde{u}$ are assumed for the parameters $p$ and $\tilde{p}\,$, respectively. The equality $u \, \equiv \, \tilde{u}$ yields to:
\begin{align}
\label{eq:equality_BET}
\frac{\mathfrak{N}_{\,1}(\,a\,)}{\mathfrak{D}_{\,3}(\,a\,)} \, \equiv \, \bold{0}\,, 
\end{align}
where $\mathfrak{N}_{\,1}$ and $\mathfrak{D}_{\,3}$ are first and third order polynomials of $a\,$, where the first is:
\begin{align*}
\mathfrak{N}_{\,1} & \egal \alpha_{\,1} \ a \plus  \alpha_{\,0} \,, \qquad
\alpha_{\,0} \egal - \, p_{\,1} \ p_{\,2} \plus \tilde{p}_{\,1} \ \tilde{p}_{\,2}  \\
\alpha_{\,1} & \egal \-\,p_{\,1} \ p_{\,2} \ \tilde{p}_{\,2}  \plus \tilde{p}_{\,1} \ p_{\,2} \ \tilde{p}_{\,2} \plus 
p_{\,1} \ p_{\,2} \moins \tilde{p}_{\,1} \ \tilde{p}_{\,2} \,.
\end{align*}
Thus, Eq.~\eqref{eq:equality_BET} is obtained if 
\begin{align}
\label{eq:equality_BET2}
\alpha_{\,1} \egal \alpha_{\,0} \, \equiv \, 0 \,.
\end{align}
Eq.~\eqref{eq:equality_BET2} leads to $p_{\,1} \, \equiv \, \tilde{p}_{\,1}$ and $p_{\,2} \, \equiv \, \tilde{p}_{\,2}$ so both parameters are SGI.

\subsection{VG model}

A first set of observable $u$ is obtained for the model with parameters $p \egal \bigl(\,p_{\,1}\,,\,p_{\,2}\,,\,p_{\,3}\,)\,$. A second one $\tilde{u}$ is also hold for the parameters $\tilde{p} \egal \bigl(\,\tilde{p}_{\,1}\,,\,\tilde{p}_{\,2}\,,\,\tilde{p}_{\,3}\,)\,$. Writing $u \moins \tilde{u} \, \equiv \, 0$ gives:
\begin{align*}
\frac{p_{\,1} \ \biggl( \, 1 \plus \Bigl(\, - \, p_{\,2} \ \ln \bigl(\, a \, \bigr)  \,\Bigr)^{\,p_{\,3}} \,\biggr)^{\,-1\plus \frac{1}{p_{\,3}}}}
{\tilde{p}_{\,1} \ \biggl( \, 1 \plus \Bigl(\, - \, \tilde{p}_{\,2} \ \ln \bigl(\, a \, \bigr)  \,\Bigr)^{\,\tilde{p}_{\,3}} \,\biggr)^{\,-1\plus \frac{1}{\tilde{p}_{\,3}}}} \, \equiv \, \bold{1}
 \,.
\end{align*}
If we assume $p_{\,1} \, \equiv \, \tilde{p}_{\,1}$ and $p_{\,3} \, \equiv \, \tilde{p}_{\,3}$, then we obtain:
\begin{align*}
\frac{1 \plus \Bigl(\, - \, p_{\,2} \ \ln \bigl(\, a \, \bigr)  \,\Bigr)^{\,p_{\,3}} }
{1 \plus \Bigl(\, - \, \tilde{p}_{\,2} \ \ln \bigl(\, a \, \bigr)  \,\Bigr)^{\,p_{\,3}}  } \, \equiv \, \bold{1}
 \,.
\end{align*}
It follows that $p_{\,2} \, \equiv \, \tilde{p}_{\,2}$ and therefore the three parameters are SGI by reciprocity. 

\subsection{SM model}

Two sets of observable are hold $u$ and $\tilde{u}$ for the \textsc{Smith} model with parameters $p \egal \bigl(\,p_{\,1}\,,\,p_{\,2}\,)$ and $\tilde{p} \egal \bigl(\,\tilde{p}_{\,1}\,,\,\tilde{p}_{\,2}\,)\,$. The equality $u \, \equiv \, \tilde{u}$ provides:
\begin{align*}
p_{\,1} \plus p_{\,2} \ \ln \bigl(\, 1 \moins a \,\bigr) \, \equiv \, 
\tilde{p}_{\,1} \plus \tilde{p}_{\,2} \ \ln \bigl(\, 1 \moins a \,\bigr) \,,
\end{align*}
which can be rewritten as:
\begin{align*}
p_{\,1} \moins \tilde{p}_{\,1} \plus \bigl(\, p_{\,2} \moins \tilde{p}_{\,2} \, \bigr) \ \ln  \bigl(\, 1 \moins a \,\bigr) \, \equiv \, \bold{0} \,.
\end{align*}
Since the therm $ \ln  \bigl(\, 1 \moins a \,\bigr)$ is linearly independent, then $p_{\,1} \, \equiv \, \tilde{p}_{\,1}$ and $p_{\,2} \, \equiv \, \tilde{p}_{\,2}\,$. So parameters $p_{\,1}$ and $p_{\,2}$ are SGI. 

\subsection{MADS model}

We assume two sets of observable $u$ and $\tilde{u}$ for the MADS model, associated to the parameters $p \egal \bigl(\,p_{\,1}\,,\,p_{\,2}\,)$ and $\tilde{p} \egal \bigl(\,\tilde{p}_{\,1}\,,\,\tilde{p}_{\,2}\,)\,$, respectively. Then, with the model definition \ref{eq:MADS_model}, $u \, \equiv \, \tilde{u}$ yields to:
\begin{align*}
\frac{\Bigl(\, \tan (\, p_{\,1} \plus p_{\,2} \ a \,) \moins \tan (\, p_{\,1} \,) \,\Bigr)
\cdot  \tilde{p}_{\,2} \cdot \Bigl(\, 1 \plus \tan (\,\tilde{p}_{\,1}\,)^{\,2}\,\Bigr) 
}{
p_{\,2} \cdot \Bigl(\, 1 \plus \tan (\,p_{\,1}\,)^{\,2}\,\Bigr)
\ \Bigl(\, \tan (\, \tilde{p}_{\,1} \plus \tilde{p}_{\,2} \ a \,) \moins \tan (\, \tilde{p}_{\,1} \,) \,\Bigr)}
\, \equiv \, \bold{1} \,. 
\end{align*}
One solution is that $p_{\,2} \, \equiv \, \tilde{p}_{\,2}$ so that parameter $p_{\,2}$ is SGI. Then, it follows that $\tan (\,p_{\,1} \, \equiv \, \tan (\, \tilde{p}_{\,1} \,)$ which one solution is $p_{\,1} \, \equiv \, \tilde{p_{\,1}}\,$. Therefore both parameters are SGI.

\end{document}